\documentclass[aps,prstab,onecolumn,superscriptaddress,showpacs]{revtex4}
\usepackage{graphicx}
\usepackage{url}
\usepackage[colorlinks,linkcolor=blue,anchorcolor=blue,citecolor=blue]{hyperref} 
\usepackage{algorithm,algorithmic}

\newcommand{\opal}{\textsc{OPAL}}

\newcommand{\opalcycl}{\textsc{OPAL-cycl}}

\renewcommand{\epsilon}{\varepsilon}

\renewcommand {\Re}{{\rm I \kern-2pt R}}

\newcommand {\RM}[1]{\mathrm{#1}}

\newcommand{\bs}[1]{\mathbf #1}

\begin{document}



\title{Beam Dynamics in High Intensity Cyclotrons Including Neighboring Bunch Effects: Model, Implementation and Application} 



\author{J. J. Yang}
\email{yangjianjun00@mails.tsinghua.edu.cn}
\affiliation{China Institute of Atomic Energy, Beijing, 102413, China}
\affiliation{Paul Scherrer Institut, Villigen, CH-5234, Switzerland}
\affiliation{Department of Engineering Physics, Tsinghua University, Beijing, 100084, China}
\author{A. Adelmann}
\email{andreas.adelmann@psi.ch}
\author{M. Humbel}
\author{M. Seidel}
\affiliation{Paul Scherrer Institut, Villigen, CH-5234, Switzerland}
\author{T. J. Zhang} 
\affiliation{China Institute of Atomic Energy, Beijing, 102413, China}

\noaffiliation

\begin{abstract}

Space charge effects, being one of the most significant collective effects, play an important role in high intensity cyclotrons. However, 
for cyclotrons with small turn separation, other existing effects are of equal importance. 
Interactions of radially neighboring bunches are also present, but their combined effects has not yet been investigated in any great detail.
In this paper, a new particle in cell based self-consistent numerical simulation model is presented for the first time.
The model covers neighboring bunch effects and is implemented in the three-dimensional object-oriented parallel code \opalcycl,
a flavor of the \opal \  framework. We discuss this model together with its implementation and validation.
Simulation results are presented from the PSI 590\,MeV Ring Cyclotron in the context of the ongoing high intensity upgrade program,
which aims to provide a beam power of 1.8\,MW (CW) at the target destination.

\end{abstract}


\maketitle

\section{INTRODUCTION \label{intro}}
Since the invention of the classic cyclotron several decades ago, increasingly higher beam intensities are required  
to provide more powerful tools for many new scientific endeavors, such as Spallation Neutron Sources
and in the future Accelerator Driven Systems that are foreseen to reduce nuclear waste.
In high intensity cyclotrons, space charge effects play an important role for the following reasons.
Firstly, with the absence of longitudinal focusing in cyclotrons, the longitudinal space charge force causes additional acceleration for 
head particles and deceleration for tail particles, which may lead to an increase in energy spread. Secondly, there is a strong 
radial-longitudinal coupling which is influenced by non-linear radial and longitudinal space charge forces. 
Lastly, the space charge force can reduce the vertical tune and increase the vertical beam size, which can cause beam losses when the vertical size 
extends beyond the aperture of the accelerator \cite{Baartman:1}.

As shown in \cite{Adam:0,Adam:1,Adam:2, Bert:2001,Ada:1,Poz:1} and experimentally verified at the PSI Injector II \cite{HumbPC}, an intense particle beam which is properly matched to a separate sector cyclotron, 
develops a spatial stationary circular bunch distribution. Consequently the halo is significantly reduced and a corresponding longitudinal decrease of the beam size is observed. 
We note that Kleeven gave in his thesis \cite{kleeven:1} an ``early'' hint on this remarkable effect and in \cite{Bert:2001} a compact derivation can be found.
On a very practical side, because of the compact stationary distribution 
we can operate the 3$^{rd}$ harmonic resonator, the former flat-top resonator in the PSI Injector II, in acceleration mode.

Non-linear space charge effects in cyclotrons are complex because of the complicated magnetic topology (reference trajectory with non constant
curvature). Typically there are two approaches 
to deal with this difficulty: one is an approximation with analytic and semi-analytic models which are inexpensive to compute \cite{Bert:2001} \cite{kleeven:1}. With such models we can obtain a qualitative  
understanding of the problem. 
In the past, a number of models have been developed based on this philosophy, such as the Sector Model, Disk Model, Sphere Model and the so called Needle Model \cite{Gordon:1,Joho:1,Adam:1,Adam:2}. 
Another more accurate approach is numerical simulation with macro-particles. In this field, typically two different technologies have used to 
solve space charge fields: Particle-Particle (P-P) methods \cite{Li:1} and Particle-Mesh (P-M) methods \cite{Ada:1,Poz:1}. In P-P methods, the fields imposed on 
a given particle are obtained by directly summing up the contributions of all other particles at this position. Limited by its low computation efficiency $\mathcal{O}(n^2)$ with $n$ denoting the number of simulation particles, 
it is impossible to utilize these methods when the number of particles is large (above 1\,million), even on current state-of-the-art supercomputers. In contrary, in P-M methods
the fields are calculated on the discrete domains. Due to its high efficiency and high precision, P-M based Particle-In-Cell (PIC) methods \cite{Hockney:1} 
are widely used in parallel macro-particles simulation codes for different types of accelerators and beam lines \cite{Ada:1,Qiang:1,Gala:1,Grote:1,Huang:1} 
as well as many other areas of computational science, thanks to the development of parallel computation technology during recent years. 
The Parallel PIC model is the method of choice in this study on the beam dynamics of high intensity cyclotrons.

For high intensity cyclotrons, single bunch space charge effects are not the only contribution. 
Along with the steady increase of beam current, the mutual interaction of neighboring bunches in radial direction 
becomes more and more important, especially at large radii where the distances between neighboring bunches diminishes, and even the overlap can occur.
One example is the PSI 590\,MeV Ring Cyclotron \cite{Mike:1} with a production current of 2\,mA in CW operation and a beam power on target of approximately 1.2\,MW.
An ambitious upgrade program for the PSI Ring Cyclotron is in progress, aimed for 1.8\,MW CW beam power on target. 
The concept involves replacing four aluminum cavities by new copper cavities with peak voltages
increasing from about 0.7\,MV to above 0.9\,MV, meanwhile the old flat-top cavity remains in use with peak voltage standing at 11.2$\%$ of the main voltage.
 After the planned upgrade is finished, the total turn number can be significantly reduced, e.g. from more than 220 turns to less than 
170 turns.

\begin{figure}
    {\includegraphics[width=8cm,trim=2.5cm 2.5cm 2.5cm 2.5cm]{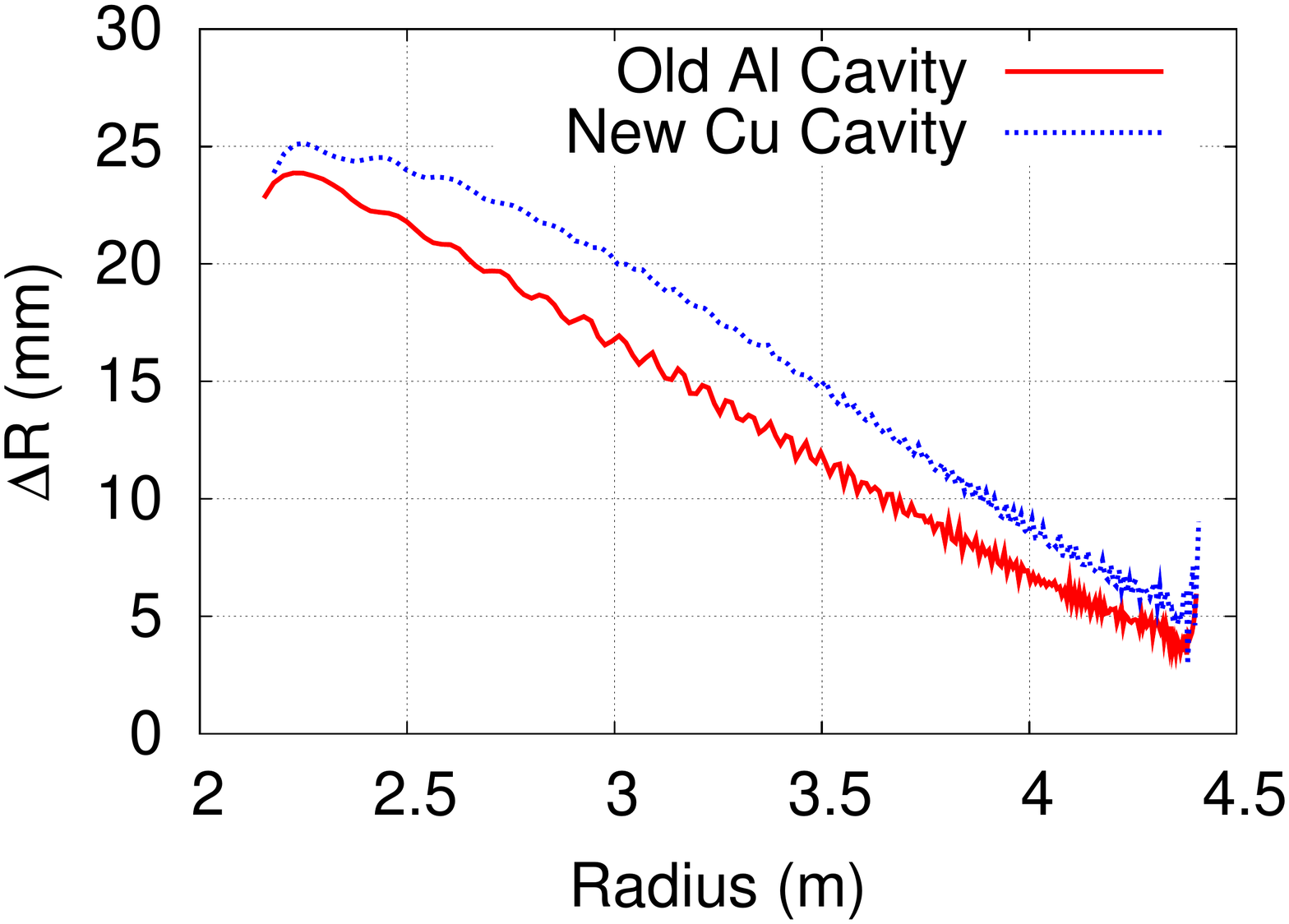}}
    {\includegraphics[width=8cm,trim=2.5cm 2.5cm 2.5cm 2.5cm]{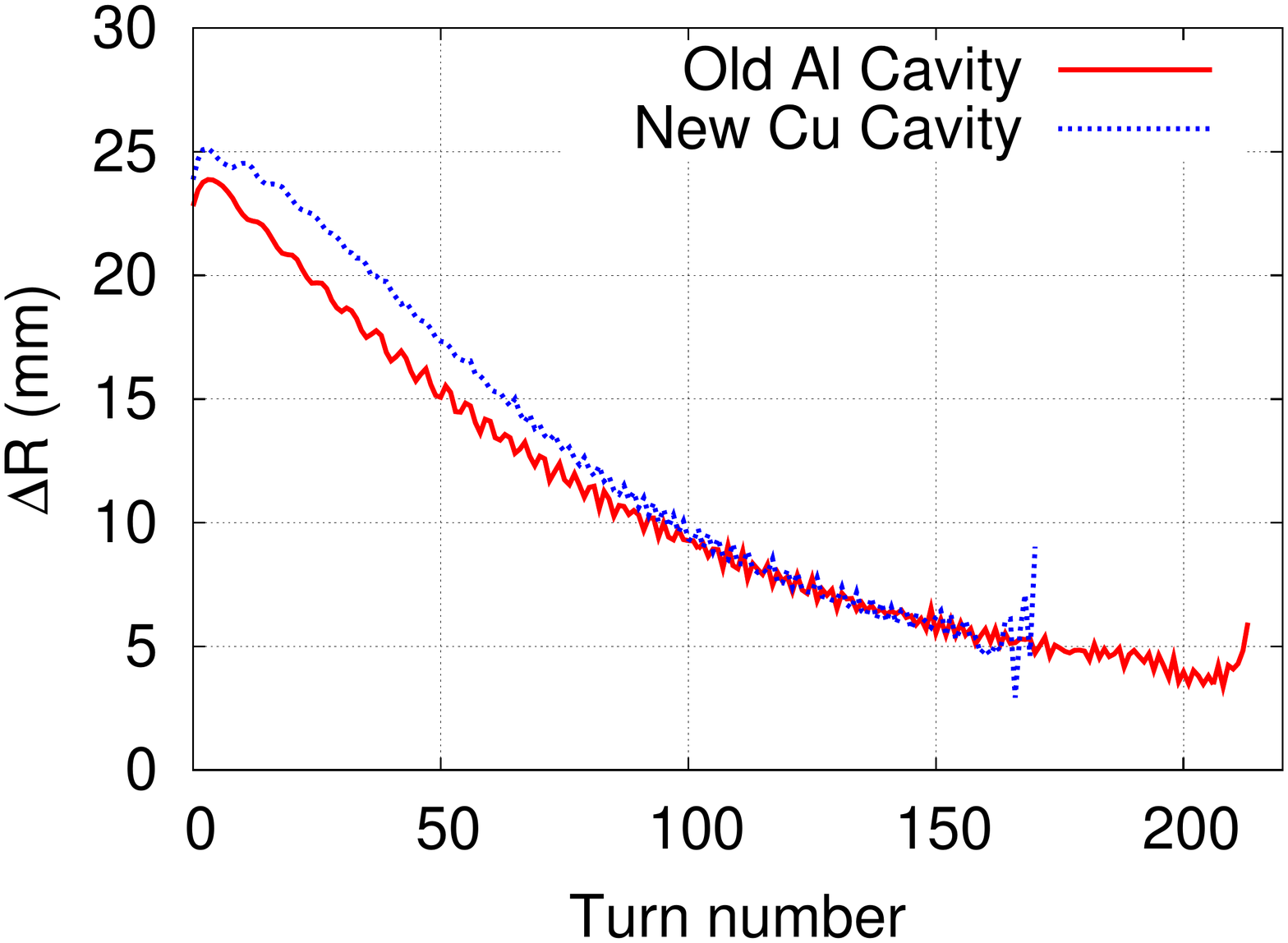}}
    \caption{(Color) Comparison of calculated turn separation for centroid particles before (red line) and after (blue line) upgrade of the PSI Ring Cyclotron.
      The main cavity voltage is set to 0.735 MV and 0.9\,MV for aluminum and copper respectively while the flat-top is set to 11.2$\%$ of the main voltage.}
    \label{fig:TuneSep}
  \end{figure}

The turn separation is consequently increased as shown in Fig.\,\ref{fig:TuneSep}, but remains at the same order of magnitude as the measured radial bunch size (at the $1\sigma$ level) and is also dependent on the increasing bunch current.
Therefore, when the beam current increases from 2\,mA to 3\,mA, the correct treatment of space charge effects is of great importance. This includes
the mutual space charge effects between radially neighboring bunches.

Another example is the 100\,MeV compact cyclotron (CYCIAE-100) under construction at CIAE \cite{Zhang:1}. Although its beam current is only 200\,$\mu$A to 500\,$\mu$A,
because of the small energy gain per turn (about 200\,keV), the turn separation is far more smaller than the beam size at outer radii (at the extraction, $\Delta R = 1.5\RM{\,mm}$) and multiple bunches will overlap.

Because of the complexity of the problem, it is impossible to evaluate neighboring bunch effects precisely and self-consistently by explicit 
analytic expressions. However, high performance computation (HPC) makes it possible to 
treat this problem in greater detail. To our knowledge, little research has been undertaken in neighboring bunch effects, 
and the only published work in that regard was by 
E.~Pozdeyev \cite{Poz:1}. He introduced ``rigid auxiliary bunches'' in his serial code CYCO which uses the azimuthal angle as the independent variable. 

In Section II, a new PIC based self-consistent numerical simulation model is presented which, for the first time, covers neighboring bunch effects. 
Section III describes our three-dimensional object-oriented parallel simulation code \opalcycl, a flavor of the \opal \  framework. 
The results of performance testing and code benchmarking are presented in Section IV, 
and followed by its applications on the PSI 590\,MeV Ring Cyclotron in Section V. 
The last section is devoted to teh conclusion of the paper.

\section{BASIC EQUATIONS AND PHYSICAL MODEL }
\subsection{PIC MODEL IN CYCLOTRON}
In the cyclotrons and beam lines under consideration, the collision between particles can be neglected because the typical bunch densities are low.
In time domain, the general equations of motion of charged particle in electromagnetic fields can be expressed by
\begin{equation}\label{eq:motion}
  \frac{d\bs{p}(t)}{dt}  = q\left(c\mbox{\boldmath$\beta$}\times \bs{B} + \bs{E}\right), \nonumber \\
\end{equation}
where $m_0, q,\gamma$ are rest mass, charge and the relativistic factor. With $\bs{p}=m_0 c \gamma \mbox{\boldmath$\beta$}$ we denote the momentum of a particle, 
$c$ is the speed of light, and $\mbox{\boldmath$\beta$}=(\beta_x, \beta_y, \beta_z)$ is the normalized velocity vector. In general the time ($t$) and position ($\bs{x}$) dependent electric and magnetic vector fields are
written in abbreviated form as $\bs{B} \text{ and } \bs{E}$.

If $\bs{p}$ is normalized by $m_0c$, 
Eq.\,(\ref{eq:motion}) can be written in Cartesian coordinates as 
\begin{eqnarray}\label{eq:motion2}
  \frac{dp_x}{dt} & = & \frac{q}{m_0c}E_x + \frac{q}{\gamma m_0}(p_y B_z - p_z B_y),    \nonumber \\
  \frac{dp_y}{dt} & = & \frac{q}{m_0c}E_y + \frac{q}{\gamma m_0}(p_z B_x - p_x B_z),   \\
  \frac{dp_z}{dt} & = & \frac{q}{m_0c}E_z + \frac{q}{\gamma m_0}(p_x B_y - p_y B_x).    \nonumber 
\end{eqnarray}
The evolution of the beam's distribution function $ f(\bs {x},c\mbox{\boldmath$\beta$},t)$ can be expressed by a collisionless Vlasov equation:
\begin{equation}\label{eq:Vlasov}
  \frac{df}{dt}=\partial_t f + c\mbox{\boldmath$\beta$} \cdot \nabla_x f +q(\bs{E}+ c\mbox{\boldmath$\beta$}\times\bs{B})\cdot \nabla_{c\mbox{\boldmath$\beta$}} f  =  0, 
\end{equation}
where $\bs{E}$ and $\bs{B}$ include both external applied fields, space charge fields and other collective effects such as wake fields
\begin{eqnarray}\label{eq:Allfield}
  \bs{E} & = & \bs{E_{\RM{ext}}}+\bs{E_{\RM{sc}}}, \nonumber\\    
  \bs{B} & = & \bs{B_{\RM{ext}}}+\bs{B_{\RM{sc}}}.
\end{eqnarray}
In order to model a cyclotron, the external electromagnetic fields are given by measurement or by numerical calculations. 

The space charge fields can be obtained
by a quasi-static approximation. In this approach, the relative motion of the particles is non-relativistic in the beam rest frame, so the self-induced magnetic field is practically absent and the electric field can be computed by solving Poisson's equation
\begin{equation}\label{eq:Poisson}
  \nabla^{2} \phi(\bs{x}) = - \frac{\rho(\bs{x})}{\varepsilon_0},
\end{equation}
where $\phi$ and $\rho$ are the electrostatic potential and the spatial charge density in the beam rest frame. The electric field can then be calculated by
\begin{equation}\label{eq:Efield}
  \bs{E}=-\nabla\phi,
\end{equation}
and back transformed to yield both the electric and the magnetic fields, in the lab frame, required in Eq.\,(\ref{eq:Allfield}) by means of a Lorentz transformation.
Because of the large gap in our cyclotron, the contribution of image charges and currents are minor effects compared to space charges \cite{Baartman:1}, and hence it is a good approximation to use 
open boundary conditions. 

The combination of Eq.\,(\ref{eq:Vlasov}) and Eq.\,(\ref{eq:Poisson}) constitutes the Vlasov-Poisson system. 
In the content followed, the method of how to solve these equations in cyclotrons using PIC methods is described in detail.

\begin{figure}
    {\includegraphics[width=8cm]{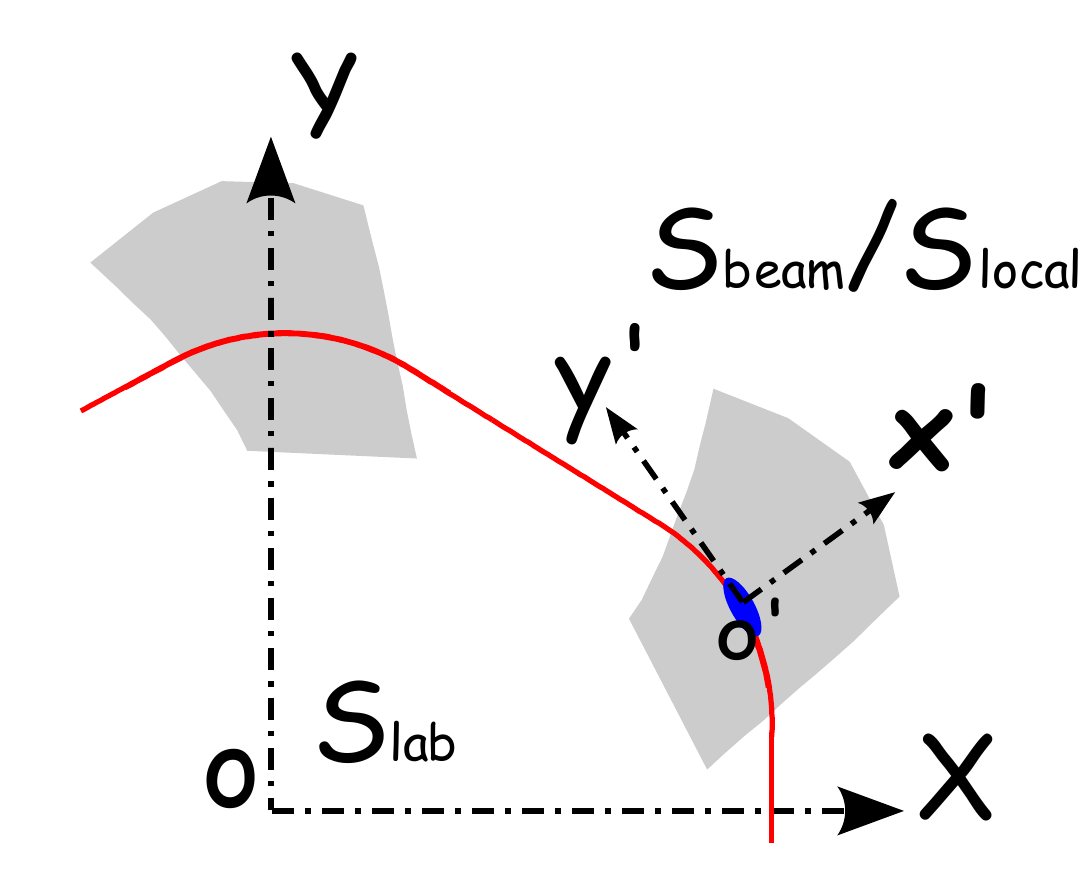}}
    \caption{(Color) Schematic plot of the top view of the three coordinates frames. The red curve is the orbit of bunch center, 
      the blue area represents bunch shape, and the gray area is the hill region of magnetic field.}
    \label{fig:frame}
  \end{figure}

Considering that particles propagates spirally outwards in cyclotrons, and the longitudinal orientation changes continuously,
three right-handed Cartesian coordinate systems are defined, as shown in Fig.\,\ref{fig:frame}.  
The first coordinate system is the fixed laboratory frame ${\bs{S}_{\RM{lab}}}$, in which the external field data is stored and the particles are tracked. 
Its origin is the center of the cyclotron and its $X-Y$ plane is the median plane and the positive direction of $Z$ axis points to vertical direction.

The second coordinate system is the local instantaneous frame ${\bs{S}_{\RM{local}}}$, which is a temporal auxiliary frame for the space charge solver.
Its origin $O'$ is the mass center of the beam and the orientation of the $Y'$ axis is coincident with the average longitudinal direction and 
the positive orientation of the $Z'$ axis points to the vertical direction.

The third coordinate system is the beam rest frame $\bs{S}_{\RM{beam}}$, which is co-moving with the centroid of the beam. 
It has the same orientation and origin as ${\bs{S}_{\RM{local}}}$, but the length in longitudinal 
direction is scaled by $1/\gamma$ due to relativistic effects. 

At each time step, in order to seek a solution for the space charge fields, the frames $\bs{S}_{\RM{local}}$ and $\bs{S}_{\RM{beam}}$ are redefined according to current 6D 
phase space distribution, and all particles are transformed from $\bs{S}_{\RM{lab}}$ to $\bs{S}_{\RM{local}}$, 
then a Lorentz transformation is performed to transform all particles to $\bs{S}_{\RM{beam}}$.
The Poisson equation is then solved in the frame $\bs{S}_{\RM{beam}}$. In a 3D Cartesian frame, the solution of the Poisson equation at point $(x,y,z)$ can be expressed by 
\begin{equation}\label{eq:Poten}
  \phi(x,y,z)= \frac{1}{4\pi\varepsilon_0}\int{G(x,x',y,y',z,z')\rho(x',y',z')dx'dy'dz'},
\end{equation}
with $G$ the 3D Green function 
\begin{equation}\label{eq:Green}
  G(x,x',y,y',z,z')= \frac{1}{\sqrt{(x-x')^2+(y-y')^2+(z-z')^2}},
\end{equation}
assuming open boundary conditions.
The typical steps of calculating space charge fields using the Hockney's FFT algorithm \cite{Hockney:1} is sketched in Algorithm \ref{alg1:sc3d},
where the quantities with superscript $D$ (discrete) refer to grid quantities.

\begin{algorithm}
  \caption{3D Space Charge Calculation} 
  \label{alg1:sc3d}
  \begin{algorithmic}[1]
    \STATE \textbf{procedure} 3DSpaceCharge(In: $\rho$, $G$, Out: $\bs{E_{sc}}$,$\bs{B_{sc}}$)
       \STATE Create 3D rectangular grid which contains all particles 
       \STATE Interpolate the charge $q$ of each macro-particle to nearby mesh points to obtain $\rho^D$
       \STATE Lorentz transformation to obtain $\rho^D$ in the beam rest frame $\bs{S}_{\RM{beam}}$
       \STATE FFT $\rho^D$ and $G^D$ to obtain $\widehat{\rho}^D$ and $\widehat{G}^D$
       \STATE Determine $\widehat{\phi}^D$ on the grid using $\widehat{\phi}^D = \widehat{\rho}^D \cdot \widehat{G}^D$
       \STATE Use FFT$^{-1}$ of $\widehat{\phi }^D$ to obtain $\phi^D$
       \STATE Compute $\bs{E}^D= -\nabla \phi^D$
       \STATE Interpolate $\bs{E}$ at the particle positions $\bs{x}$ from $\bs{E}^D$
       \STATE Inverse Lorentz transform to obtain $\bs{E_{\RM{sc}}}$ and $\bs{B_{\RM{sc}}}$ in  frame $\bs{S}_{\RM{local}}$ and transform back  to $\bs{S}_{\RM{lab}}$
       \STATE \textbf{end procedure}
  \end{algorithmic}
\end{algorithm}
With respect to the external magnetic field two possible situations can be considered: 
in the first situation, the real field map is available on the median plane of the existing cyclotron machine using measurement equipment.
In most cases concerning cyclotrons, the vertical field, $B_z$, is measured on the median plane ($z=0$) only.
Since the magnetic field outside the median plane is required to compute trajectories with $z \neq 0$, the field needs to be expanded in the $Z$ direction. 
According to the approach given by Gordon and Taivassalo \cite{Gordon:2}, by using a magnetic potential and measured $B_z$ on the median plane
at the point $(r,\theta, z)$ in cylindrical polar coordinates, the 3$rd$ order field can be written as    
\begin{eqnarray}\label{eq:Bfield}
  B_r(r,\theta, z) & = & z\frac{\partial B_z}{\partial r}-\frac{1}{6}z^3 C_r, \nonumber\\    
  B_\theta(r,\theta, z) & = & \frac{z}{r}\frac{\partial B_z}{\partial \theta}-\frac{1}{6}\frac{z^3}{r} C_{\theta}, \\     
  B_z(r,\theta, z) & = & B_z-\frac{1}{2}z^2 C_z,  \nonumber    
\end{eqnarray}
where $B_z\equiv B_z(r, \theta, 0)$ and  
\begin{eqnarray}\label{eq:Bcoeff}
  C_r & = & \frac{\partial^3B_z}{\partial r^3} + \frac{1}{r}\frac{\partial^2 B_z}{\partial r^2} - \frac{1}{r^2}\frac{\partial B_z}{\partial r} 
        + \frac{1}{r^2}\frac{\partial^3 B_z}{\partial r \partial \theta^2} - 2\frac{1}{r^3}\frac{\partial^2 B_z}{\partial \theta^2}, \nonumber  \\    
  C_{\theta} & = & \frac{1}{r}\frac{\partial^2 B_z}{\partial r \partial \theta} + \frac{\partial^3 B_z}{\partial r^2 \partial \theta}
        + \frac{1}{r^2}\frac{\partial^3 B_z}{\partial \theta^3},  \\
  C_z & = & \frac{1}{r}\frac{\partial B_z}{\partial r} + \frac{\partial^2 B_z}{\partial r^2} + \frac{1}{r^2}\frac{\partial^2 B_z}{\partial \theta^2}. \nonumber
\end{eqnarray}

All the partial differential coefficients are computed on the median plane data by interpolation, using Lagrange's 5-point formula.

In the other situation, 3D field for the region of interest is calculated numerically by building a 3D model using commercial software 
during the design phase of a new cyclotron. In this case the calculated field will be more accurate, especially at large distances from the median plane i.e. a
full 3D field map can be calculated. For all calculations in this paper, we use the method by Gordon and Taivassalo \cite{Gordon:2}.

Finally both the external fields and space charge fields are used to track particles for one time step using a 4$th$ order Runge-Kutta (RK) integrator, in which 
the fields are evaluated for four times in each time step. Space charge fields are assumed to be constant during one time step,
because their variation is typically much slower than that of external fields. 
    
\subsection{NEIGHBORING BUNCH EFFECTS} \label{sec:nbe}
The code is intended to model steady state conditions for the multi-bunch beam dynamics.
In cyclotrons the pattern of turn separation $\Delta R$ is affected by many factors. These include machine characteristics such as the magnetic field, 
the acceleration voltage profile, the accelerating phase of the RF resonators  and initial centering conditions of the injected bunches. 
Generally, in cyclotrons, $\Delta R$ reduces gradually  with increasing beam energy.
For machines like the PSI Injector\,II, $\Delta R$ stays sufficiently large from injection to extraction, and in such cases,  neighboring bunch effects are negligible. 
For others, like the PSI 590\,MeV Ring Cyclotron under consideration in this section, $\Delta R$ decreases strongly during the course of acceleration, 
which results in the need of considering the neighboring bunch effects in order to obtain a precise description of the beam dynamics.
In our model, we apply an iterative procedure to determine the number of bunches necessary for a converged simulation. 
Initially a single bunch with phase space density $f_0$ and the average radial position $R_s$ is injected.  
This single bunch is tracked with space charge for one revolution period $T_{r}$. 
Then the new average radial position of the bunch $R_e$ and the bunch rms size $r_{\RM{rms}}=\sqrt{x_{\RM{rms}}^{2}+y_{\RM{rms}}^2}$ 
are calculated from the actual particle distribution. The turn separation $\Delta R$ at injection position is then given by $\Delta R = R_e - R_s$. If the condition:
\begin{equation}\label{eq:dR}
  \Delta R  \le M \times r_{\RM{rms}},
\end{equation}
is fulfilled (where $M$ is a parameter given by the user), the 6D phase space is stored as $f_{R_{e}}$. 
The code is switched to multi-bunch mode, and  $f_{R_{e}}$ will be used as the initial phase space for the following $(N_B-1)$
neighbouring bunches which will be injected one by one per $T_{r}$ time, 
where $N_B$ is the number of neighboring bunches given by the user. 

If the condition of Eq.(\ref{eq:dR}) is not fulfilled, the value of $R_e$ is assigned to the variable $R_s$. 
This single bunch is tracked with space charge for another revolution period $T_{r}$, and the new average radial position of the bunch $R_e$, 
the bunch rms size $r_{\RM{rms}}$ and the turn separation $\Delta R$ are calculated again. 
After that, the condition in Eq.(\ref{eq:dR}) is evaluated and the same procedure is repeated accordingly.

The underlying assumption for this Ansatz is that all bunches have the same phase space distribution when they reach a certain position, 
i.e. when $f_{R_{e}}$ is saved.
This is realistic and reasonable when the machine is running in a steady state. 
It need be mentioned that up to that position the coherent instabilities which might be caused by neighboring bunch effects, 
are not covered by this model.

This procedure is summarized in Algorithm~\ref{alg:mbi}.

\begin{algorithm}
  \caption{Multi-Bunch Injection Algorithm} 
  \label{alg:mbi}
  \begin{algorithmic}[1]
    \STATE \textbf{procedure} Injection(In: $N_B$, $M$, $f_0$, $R_s$) 
    \STATE Inject $f_0$ at radius $R_s$, total bunch number $i_{NB}=1$
    \STATE Track $f_0$ for one revolution period $T_{r}$, obtain $f_{R_{e}}$
    \STATE Calculate radius $R_e$ and bunch size $r_{\RM{rms}}$
    \WHILE {($R_e-R_s > M \times r_{\RM{rms}}$)}
    \STATE     Save $ R_e \rightarrow R_s$
    \STATE     Track $f_{R_{e}}$ for $T_{r}$, re-calculate $R_e$ and $r_{\RM{rms}}$
    \ENDWHILE
    \STATE Save $f_{R_{e}}$
    \WHILE {($i_{NB} < N_B$)}
    \STATE Inject $f_{R_{e}} \text{ and increment } i_{NB}$
    \STATE Track all bunches for $T_{r}$
    \ENDWHILE
    \STATE \textbf{end procedure}
  \end{algorithmic}
\end{algorithm}

In the multi-bunch algorithm above, two parameters $M$ and $N_B$ are introduced to set the time of injecting new bunches and the total bunch number respectively. 
The proper settings of these two parameters are crucial for the precise  evaluation of neighboring bunches effects. 
\begin{figure}
    {\includegraphics[width=8cm]{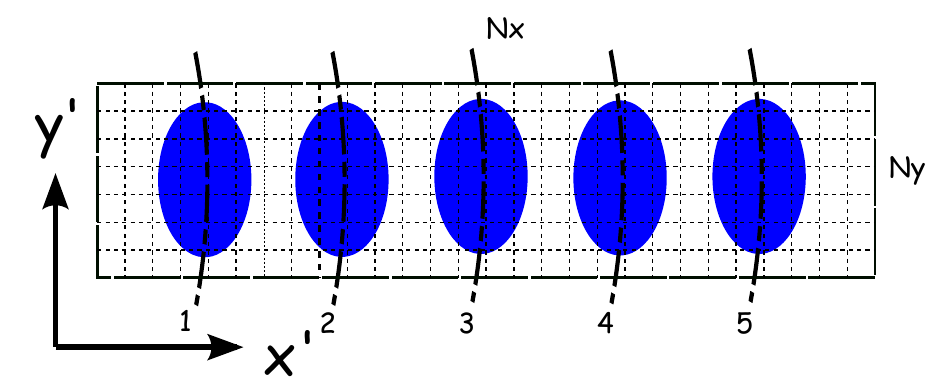}}
    \caption{(Color) Schematic plot of the top view of 5 bunches and the grid of computation domain. The grid size on $X'-Y'$ plane is Nx$\times$Ny, and the broken lines represent the orbits of bunch centers. }
    \label{fig:MultiBunch}
\end{figure}

\begin{figure}
    \includegraphics[width=9cm]{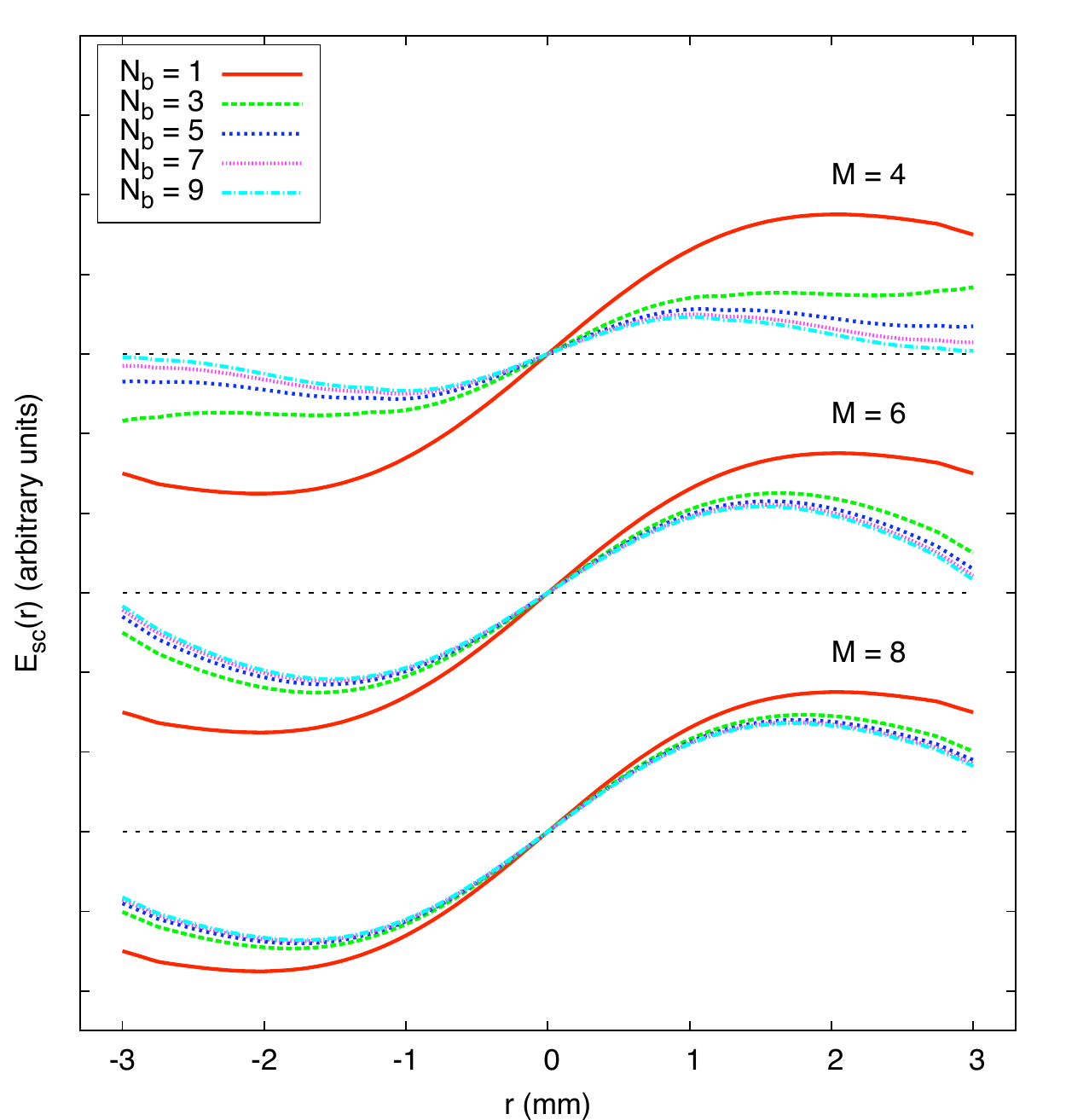}
    \caption{(Color)  The radial space charge field $E_{\RM{sc}}{(r)}$ of the multiple Gaussian beams at the central beam region.
      The beams are truncated at 10\% of their charge density and rms beam size is $r_{rms} = 1.0$ (mm). All graphs are on the same scale.}
    \label{fig:EscMB}
\end{figure}

In order to quantify the range of the two parameter $N_B$ and $M$, let's consider a 2D non-relativistic DC beam. 
The Bassetti-Erskine \cite{bassersk} formula for the electric field of a 2D Gaussian charge distribution is in general an analytical expression 
in terms of the complex error function.  In case of an axisymmetric and Gaussian charge distribution the electric field can be expressed by 
\begin{equation}\label{eq:Esc}
  \bs{E_{\RM{sc}}(r)} = \frac{I_0}{2\pi\varepsilon_0\beta cr} \left\{
    \begin{array}{ll}
      \frac{1-e^{-\frac{r^2}{2\sigma^2}}}{1-e^{-\frac{a^2}{2\sigma^2}}}\bs{n}_r, & r \le a \\
      \bs{n}_r, & r > a .
    \end{array}
    \right.
\end{equation}  
In this expression, $a$ is the truncated radius, $I_0$ the beam current and $\bs{n}_r$ unit vector in radial direction.
Using Eq.(\ref{eq:Esc}), it is easy to calculate the electrostatic field generated by $N_B$ Gaussian beams (with $N_B$ an odd number) which are all on a straight line.
We can now study the effect of $N_B$ and $M$ on the centre beam $N_{Bc} = (N_B+1)/2$ of the configuration as shown for several configurations in Fig.\,\ref{fig:EscMB}.

Although in the cyclotron, the situation is much more complex (the charge distribution and radial position of the bunches are time dependent quantities), we use this model to obtain approximate initial conditions for $N_B$ and $M$ and hence can estimate the contributions for bunches away from $N_{Bc}$.

For instance, the setting with $N_B=9$ and  $M=4.5$ gives convergent results for the PSI Ring Cyclotron with 3\,mA beam current.

In theory, when the maximum bunch number $N_B$ equals to the total turn number of the machine, one can eventually obtain the fully self-consistent solution of the problem within our model. 
In reality, it is not feasible  to simulate a full set of bunches, which typically range from 
several ten to several hundred.
The scale of the number of particles and the dimensions of the needed grid are still beyond the capability of today's supercomputer resources.


In a multi-bunch simulation the energy of the bunches at different turns can be substantially  different. 
For a multi-bunch simulation with 9 neighbouring bunches, if the kinetic energy of the first bunch is $E_k=100~\text{MeV}$ and energy gain per turn is $\Delta E_k=2~\text{MeV}$, hence
the velocity difference of the first and last bunch is $6.5\%$.  
In this case there is no single rest frame in which the relative motions of 
particles are non-relativistic, as required by our scheme to calculate the space charge forces. Consequently it is not sufficient to use only one rest frame 
and one single Lorentz transformation. In order to calculate the space charge fields more precisely, 
we use an adaptive binning technique outlined in Ref. \cite{Fubiani:2006p305} (section IV). 
We note that in our application neither radiation nor retardation effects play a significant role and can therefore be neglected. 
In the rest frame of the beam, transverse currents effects can also be neglected and hence no longitudinal magnetic field component must be considered.

First we create the same amount of energy bins ($N_B$) as we have bunches in the simulation. 
An average relativistic factor $\bar{\gamma_i}$ for the $i^{\rm th}$ bunch with $N_p^i$ simulation particles is computed,
\begin{equation}\label{eq:dR1}
  \bar{\gamma}_i = \frac{\sum_{j=1}^{N_p^i}\sqrt{1+p_{j,x}^2+p_{j,y}^2+p_{j,z}^2}}{N_p^i}, i=1\dots N_B.
\end{equation}
Then every particle is grouped into the energy bin whose $\bar{\gamma_i}$ is closest to its $\gamma$.
In this way, the energy spread of each bin is small and the relative motions of the particles in the same bin are 
small.
After binning we perform the Lorentz transformation, calculate the space charge field and perform back-transformation for each bin respectively. 
Finally the field data is summed up to give  the total space charge force imposed on each particle.

The energy spread of the bunches can be large (MeV range), especially in cyclotrons without flat-top cavities, and at large radii. This may result in an overlap of 
energy distributions of neighbouring bunches, and hence the energy bins have to be recalculated i.e.\ all particles need to be regrouped after each revolution.
It is worth noting that, in cyclotrons, the energy difference of neighboring bunches changes  with the increasing radius.
Therefore the energy difference of the neighboring bins is not constant. 
Specifically, the energy difference between the $i^{\rm th}$  and $(i-1)^{\rm th}$ bins, $\Delta\bar{E}_{i,i-1}$, 
differs with the energy difference between the $(i+1)^{\rm th}$ and the $i^{\rm th}$ bins $\Delta\bar{E}_{i+1,i}$.

\section{IMPLEMENTATION WITHIN THE \opal \  FRAMEWORK}


The above model and algorithm are implemented in the object-oriented parallel PIC code \opalcycl. 
\opalcycl \  is one of the flavors of the \opal \ (Object Oriented Parallel Accelerator Library) framework \cite{opal:1}. 
This framework is a powerful tool for charged-particle optics in general accelerator structures and beam lines using the MAD languages with extensions.
\opal ~ is  based on the CLASSIC \cite{Classic:1} library and the IP$^2$L framework \cite{ippl:1}. The CLASSIC library is a C++ class library which provides services for building portable accelerator models and algorithms and inputing 
language to specify complicated accelerator systems in general. IP$^2$L is an Object-Oriented C++ class library which provides abstractions for particles and structured field calculation
in a data-parallel programming style. It provides an integrated, layered system of objects. The upper layers
contain global data objects of physical/mathematical quantities, such as particles, fields and matrices of meshes and typical methods
performed on these objects such as differential operators and multi dimensional FFT's. The lower layers contain the objects relevant to parallelization such as data distribution, domain decomposition, communication among processors, load balancing and expression templates. Statistical data, such as root mean square (rms) quantities, are computed on the fly (in situ) and stored in conjunction with phase space and
field data in the H5Part \cite{H5part:1} file format. In a post processing step the data  can be analyzed
using the visualization tool H5PartROOT \cite{h5partroot:1}.

In addition, apart from the multi-particle simulation mode, \opalcycl \  also has two other serial tracking modes for conventional cyclotron machine design. 
One mode is the single particle tracking mode, which is a useful tool for 
the preliminary design of a new cyclotron. It allows to compute  basic parameters, such as reference orbit, phase shift history,
stable region and matching phase ellipse. The other one is the tune calculation mode, which can be used to compute the betatron oscillation frequency 
$\nu_r$ and $\nu_z$. This is useful for evaluating 
the focusing characteristics of a given magnetic field map. 

A more detailed description of the hierarchical layout, the parallelization and the implementation issues of the \opal \  framework and \opalcycl \  code
can be found in the User's Reference Guide \cite{opal:1}.

\section{PERFORMANCE TEST AND VALIDATION}
In order to evaluate the performance and to benchmark the functionalities of the newly developed code, we performed different types of simulations on
 the 72\,MeV Injector\,II cyclotron of PSI, which has been intensively studied before. Some results are presented in this section. 

\subsection{Single particle tracking and tune calculation}
\begin{figure}
  {\includegraphics[width=8cm,trim=2.5cm 2.5cm 2.5cm 2.5cm]{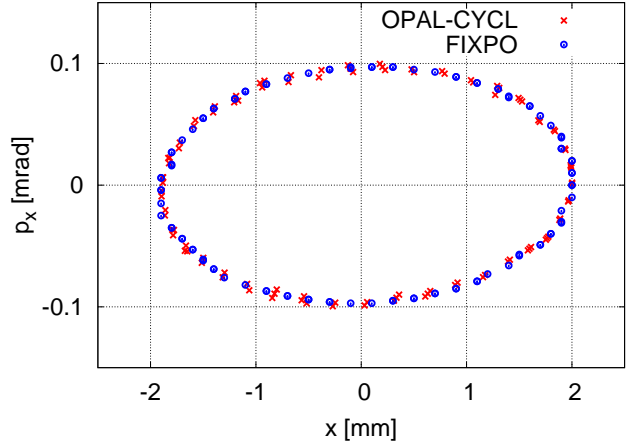}}
  \caption{(Color) Radial eigen-ellipse of 2\,MeV at the symmetric line of a magnet in Injector\,II cyclotron.}
  \label{fig:Eigen}
\end{figure}

In theory, there is an eigen-ellipse for any given energy in a cyclotron under stable conditions. When the initial phase space
matches this eigen-ellipse, the oscillation of the beam envelope amplitude will be minimal and the transmission efficiency will be maximal.
In the present test, the eigen-ellipse at 2\,MeV kinetic energy is calculated using the single particle tracking mode of \opalcycl.  
The result is compared to FIXPO \cite{FIXPO:1,joho:2}.  
At PSI the FIXPO code has been the standard simulation tool for the 
design and parameter optimisation of the Injector II and the Ring Cyclotron as well as 
for a gas driven muon trap in a cyclotron shaped magnetic field.
The code integrates single particle orbits by use of a predictor corrector algorithm up to the third order.
In Fig.\,\ref{fig:Eigen} the matched radial ellipse with an initial offset of $x=2.0{\RM{\,mm}}$, $p_x=0.0{\RM{\,mrad}}$ at the symmetry line of the sector field is shown.
Excellent agreement is obtained when the time step is set to 1\,ps in \opalcycl, although FIXPO uses a different tracking algorithm with the azimuthal angle as
the independent variable.

\begin{figure}
  {\includegraphics[width=8cm,trim=2.5cm 2.5cm 2.5cm 2.5cm]{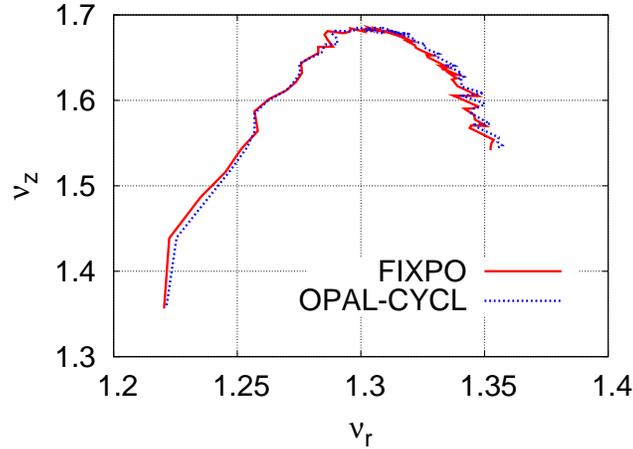}}
  \caption{(Color) Tune diagram of Injector\,II cyclotron, compared with FIXPO code.}
  \label{fig:nurnuz_Inj2}
\end{figure}

The tune diagram of Injector\,II is computed using the tune calculation mode of \opalcycl, as shown in Fig.\,\ref{fig:nurnuz_Inj2}.
The result from FIXPO and \opalcycl \ are again in good agreement even though different numerical algorithms are used.

The field interpolation scheme, particle tracking and tuning calculation functionalities are validated substantially by the above tests. 
\subsection{Parallel scalability test}
\begin{figure}
  {\includegraphics[width=8cm,trim=0cm 0cm 0cm 0cm]{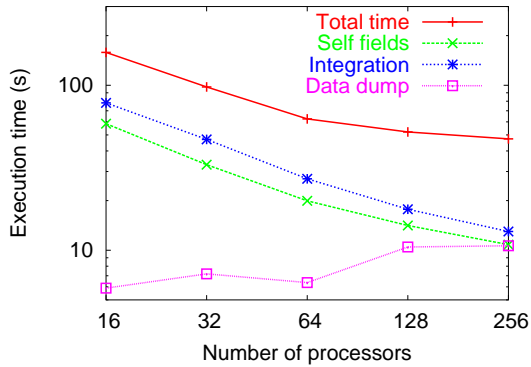}}
  \caption{(Color) The time consumption as a function of processors on Cray XT3, CSCS.}
  \label{scalability}
\end{figure}

In order to observe the parallel scalability of the code, we have performed a detailed study of strong scaling, i.e. the problem size remains constant while increasing the number of 
computing resources.
One million particles are used and tracked 200 time steps on the Injector\,II. The initial beam has a Gaussian
type distribution. The grid size is $64 \times 64 \times 64$ which is decomposed onto a two dimensional grid of processors. All the intermediate phase space data is dumped into 
a single H5Part file. The dynamic load balancing frequency as well as the phase space dumping frequency are set to 10.
The results are shown in Fig.\,\ref{scalability}.

We can see that good scalability is achieved up to 128 processors.  Above 128 processors, the time consumption of the phase space dumping starts to become significant.
The reason for the behavior is the increasing overhead in communication with respect to the amount of data to be stored 
Nevertheless, the scalability of the space charge solver and the particles integrator still benefit from a large number of processors.

\subsection{Stationary Round Distribution in the PSI Injector\,II}
Space charge effects usually result in an increase in beam size and emittance. That is detrimental to beam dynamics. 
However there are cases, where space charge effects can actually play a positive role. The PSI Injector\,II cyclotron is a space charge dominated machine, in which a very compact 
stable beam is developed within the first several turns, and thereafter, the charge distribution does not change significantly.
This stationary situation remains essentially unchanged until extraction and the beam phase width  is about $2^\circ$ in the last turn.
This is due to the combined effect of the strong coupling between the
radial and longitudinal directions in the cyclotron and the space charge when the beam current increases above 1\,mA. S.\,Koscielniak and S.\,Adam reproduced this phenomenon by 
using the serial two-dimensional code PICN \cite{Adam:2}. PICN is based on the Needle model, which treats the beam as 
an ensemble of charged vertical needles with the same height as the beam. In this model, the vertical motion of particles is separated from the horizontal 
component and the internal motion within needles is neglected. In order to validate the space charge solver module of our code, we performed the 1\,mA, 3\,MeV coasting beam simulation on this machine and compared the result with PICN.  

\begin{figure*}
    \includegraphics[angle=90,width=0.3\linewidth]{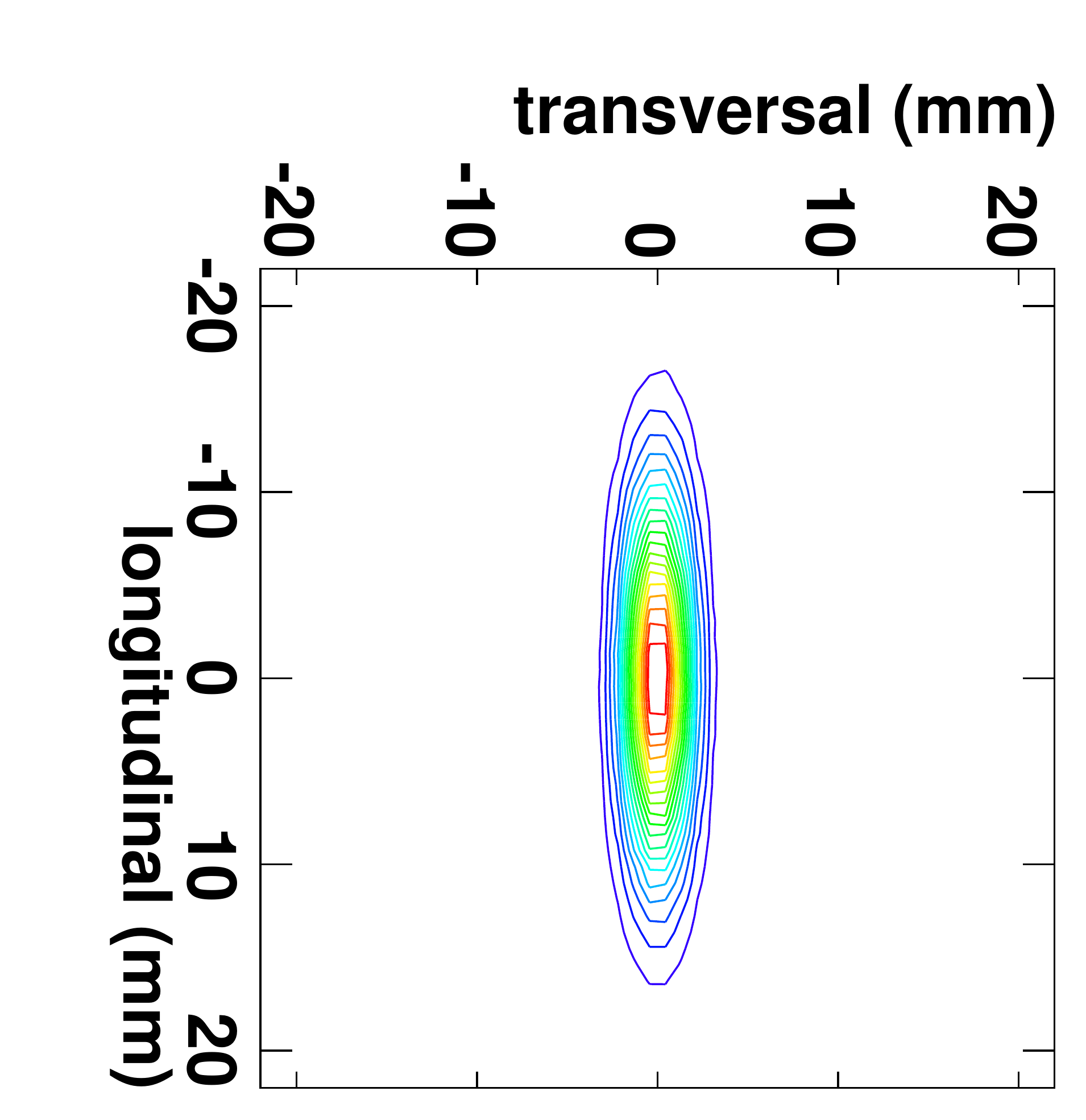}
    \includegraphics[angle=90,width=0.3\linewidth]{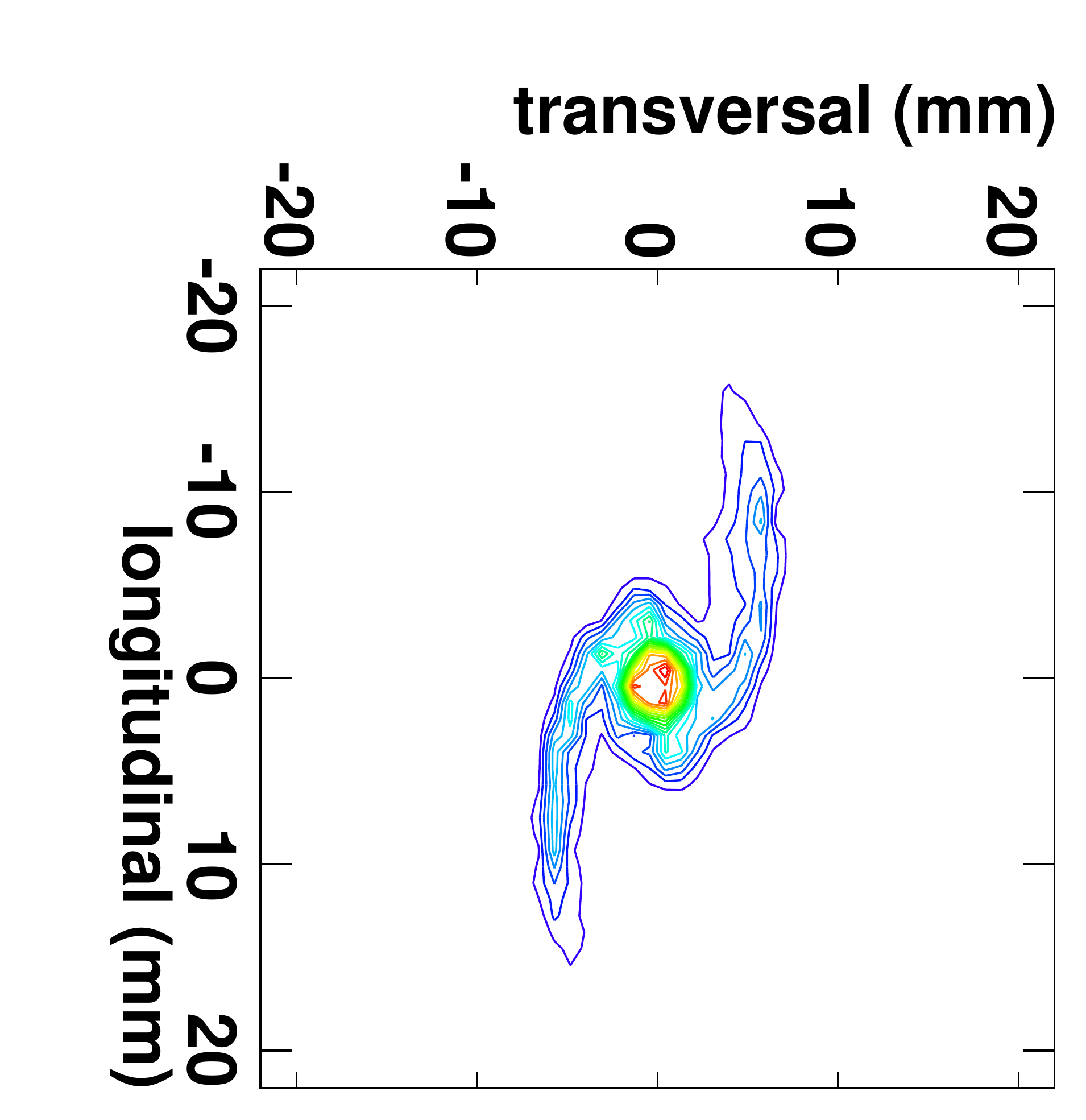}
    \includegraphics[angle=90,width=0.3\linewidth]{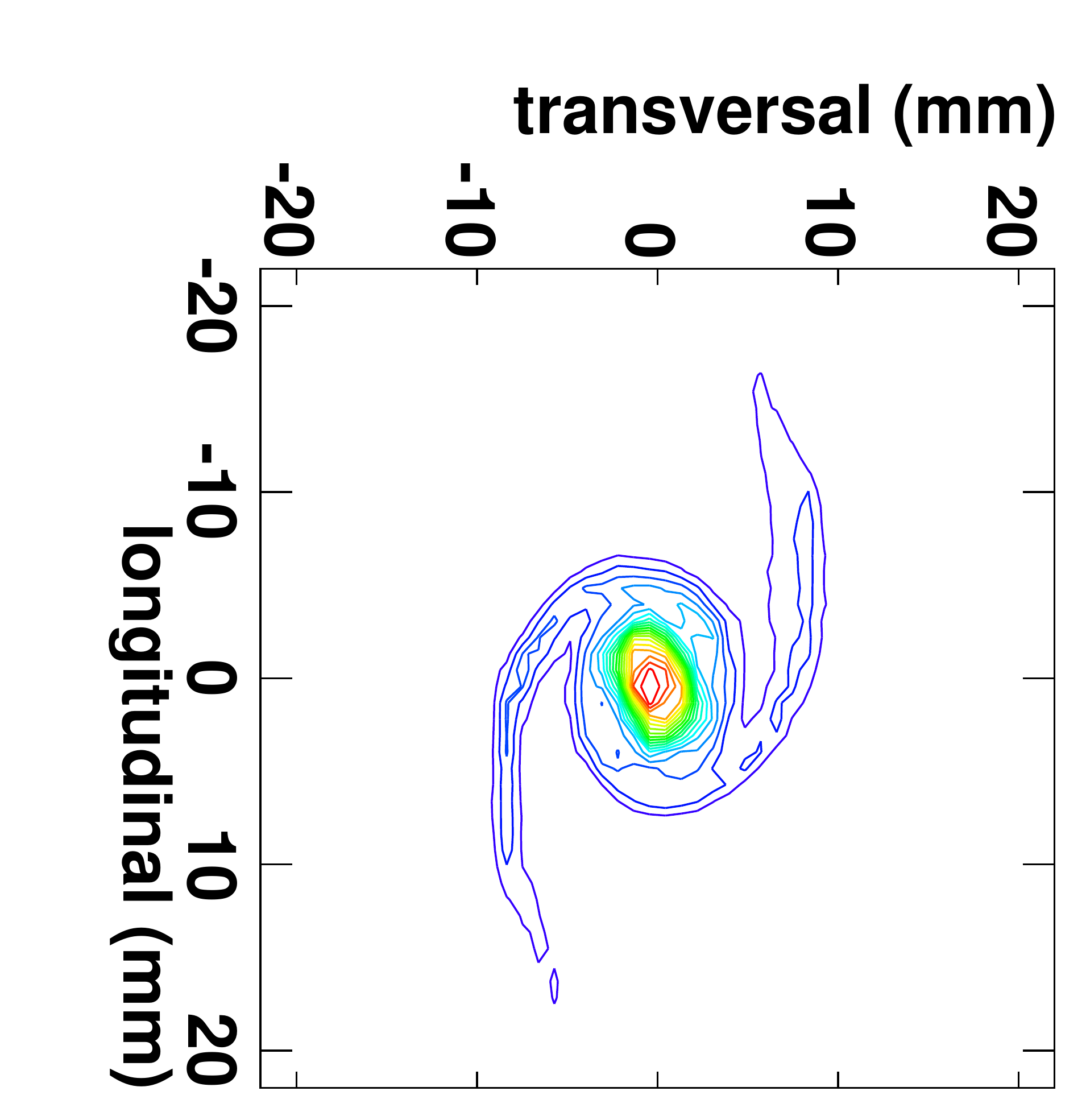}
    \includegraphics[angle=90,width=0.3\linewidth]{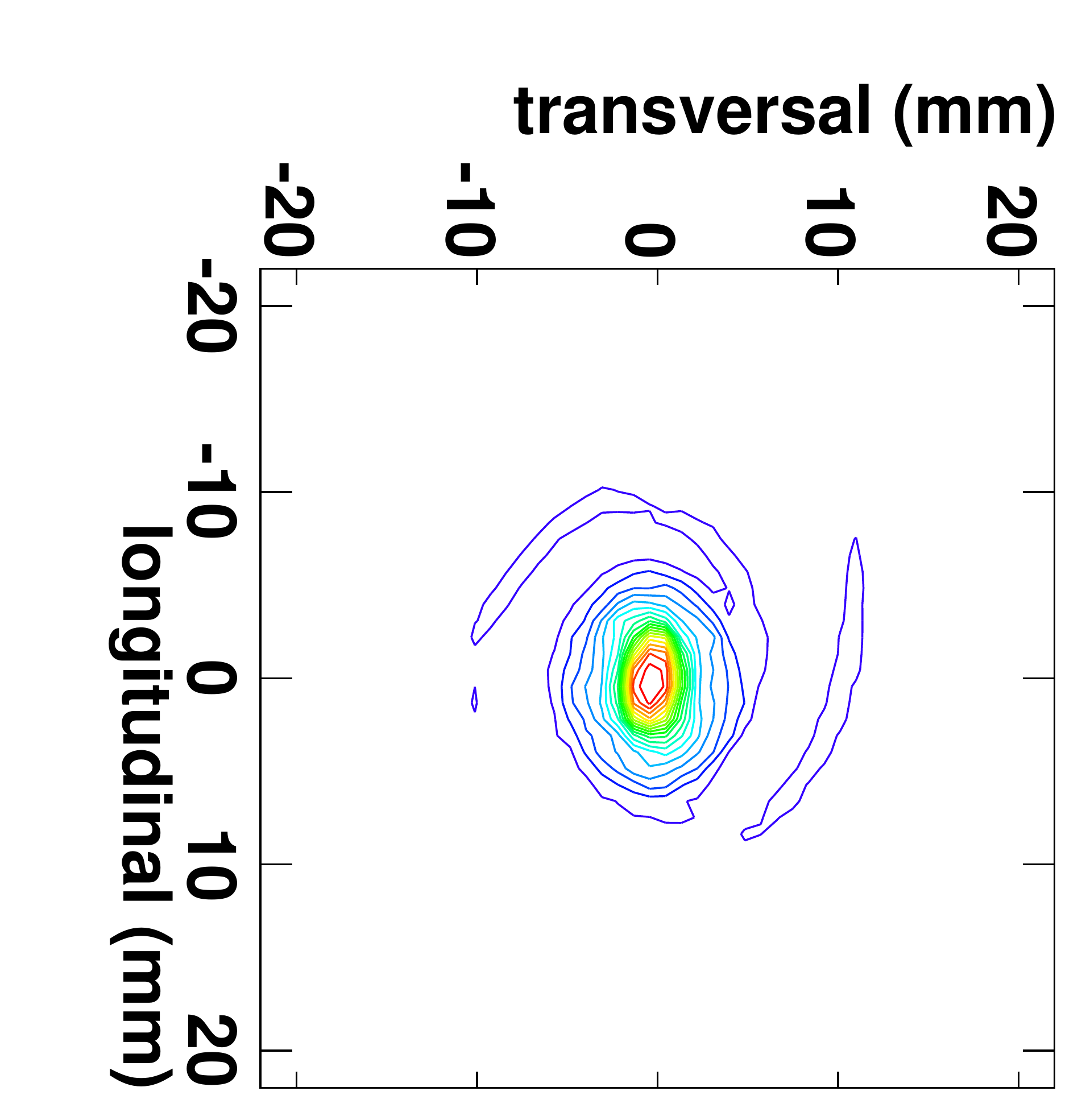}
    \includegraphics[angle=90,width=0.3\linewidth]{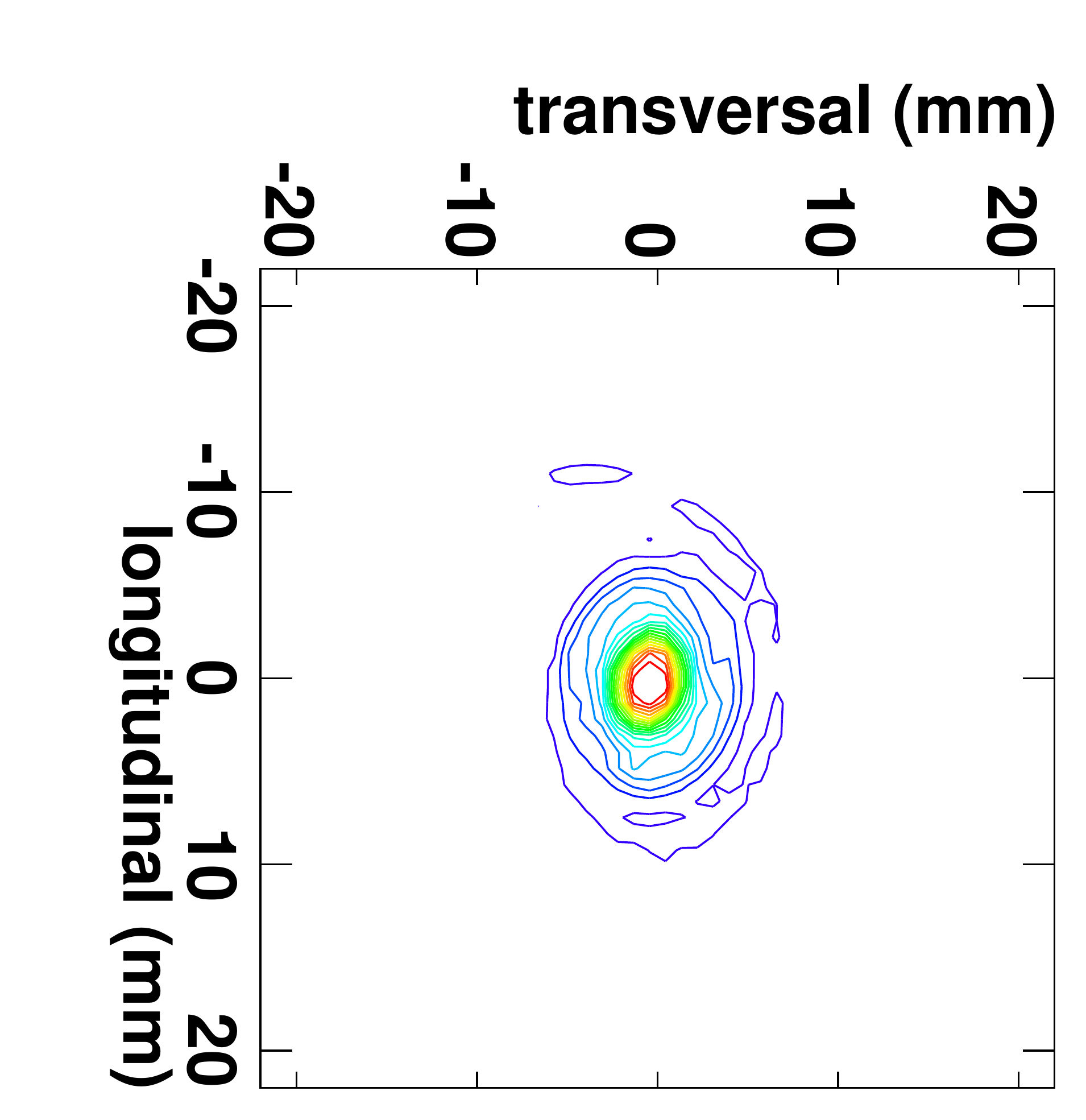}
    \includegraphics[angle=90,width=0.3\linewidth]{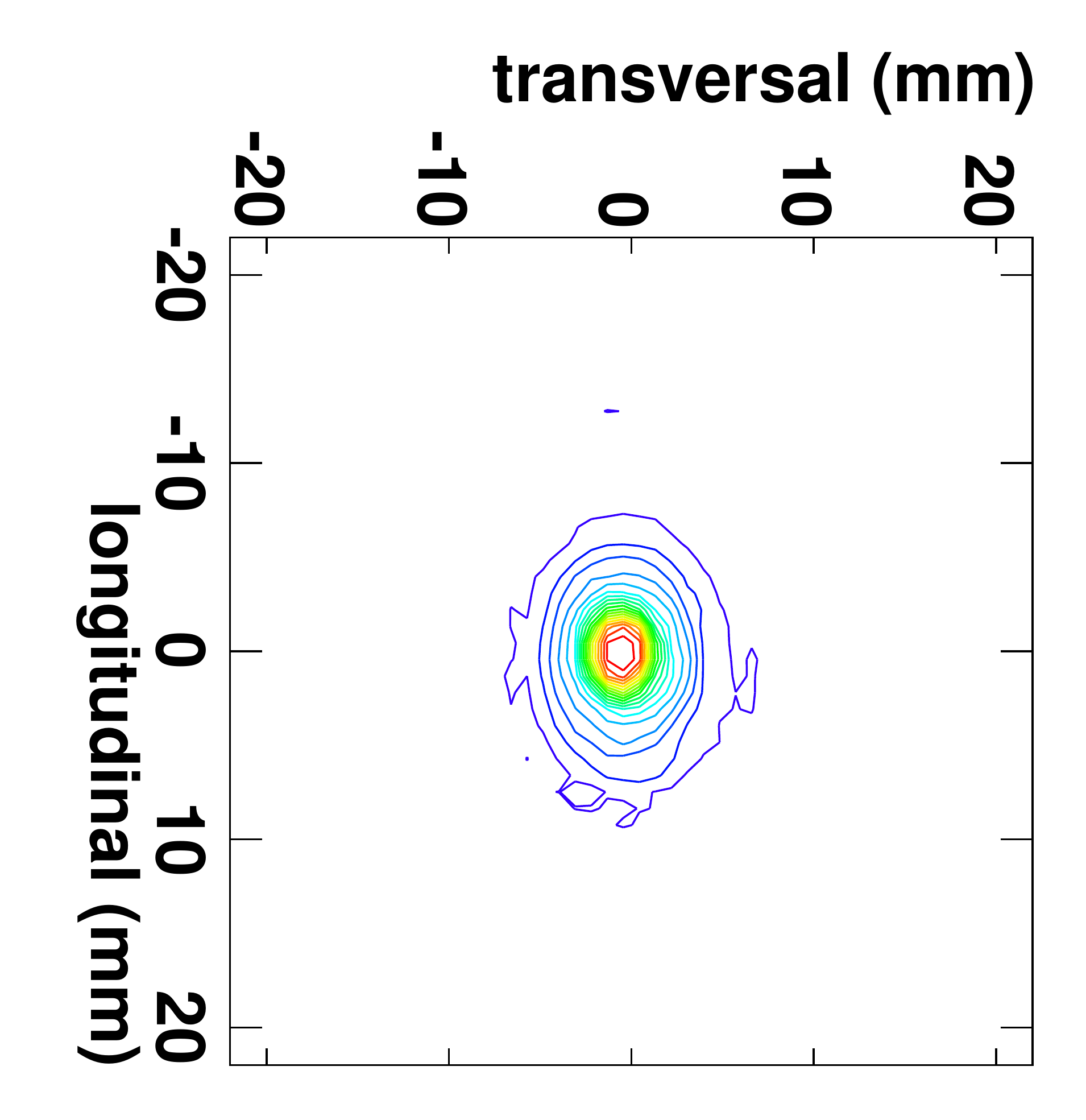}
    \caption{(Color) Top view of a 1\,mA, 3\,MeV coasting beam in PSI Injector\,II in the local frame ${\bs{S}_{\RM{local}}}$ of PSI Injector\,II. 
      Up: turn 0, 5, 10. Down: turn 20, 30, 40. To compare with figures in Ref. \onlinecite{Adam:3}, 
      the beam's transport direction is along the negative direction of the abscissa axis.}
    \label{fig:coasting1mAA}
\end{figure*}

We used the same initial distribution as in Ref. \onlinecite{Adam:3}.
$2\sigma_{\RM{longitudinal}} = 13.4 {\RM{\,mm}}$ ( $15^\circ$ phase width), $2\sigma_{\RM{transverse}} = 2.52 {\RM{\,mm}}$. 
The initial emittances of the radial and azimuthal directions are set to zero. 
In the vertical direction, $2\sigma_z = 4{\RM{\,mm}}$ and $2\sigma_{p_z} = 3.68{\RM{\,mrad}}$. 
The total macro-particles number is 1\,million.  Figure\,\ref{fig:coasting1mAA} shows the top view of beam shape in the local beam frame.
We can see that a stable core is developed after about 10 turns, which is faster than the formation of stable haloes.

When comparing Fig.\,\ref{fig:coasting1mAA} with Fig.\,3 and Fig.\,4 in Ref. \onlinecite{Adam:3} calculated by PICN, we can find the results agree with each 
other qualitatively. Both of these two codes predict the formation of a compact stable core and  wide haloes after 40 turns.
However, there still exist visible differences. PICN shows an almost ``round'' charge density distribution, in the case of \opalcycl\,  we still see low density halo. The longitudinal size of the core is about 2\,mm longer than the transverse size. This difference mainly comes from different physical models used in the two codes. 
PICN uses a so called smooth approximation which treats the particle orbits
as pure circles, but with a sinusoidal radial/vertical focusing(betatran oscillation) having a realistic value of about $1.14$ at 3 MeV. 
Of course, no realistic magnetic field can be azimuthally constant and in the same time focusing in both directions\cite{adapconv}. 
In addition all particles in close proximity to the horizontal plane are represented by a single ``needle''; in \opalcycl\  only those particles close to each other in configuration space are represented by a single macro-particle. The
latter is believed to be more realistic and accurate. 

It can be concluded from this comparison that \opalcycl\ can reliably reproduce the stable ``round'' beam formation caused by space-charge effects in Injector\,II and hence can accurately  reproduce the single bunch dynamics in cyclotrons.

\section{APPLICATIONS}
We start this section by describing the two cyclotrons under consideration. The key parameters of the machines are given in Table \ref{tab:cycs}. 
In addition we mention the fundamental RF frequency is 50.633 MHz.
In the Injector\,II due to the circular bunch formation, discussed in section IV, the original designed third harmonic flat-top resonator is now being used as an additional 
accelerating structure.
It is now obvious to ask the question if one can find a feasible working point for the Ring Cyclotron with the same characteristics as obtained in the Injector\,II. However because of
the overlapping tuns in the Ring Cyclotron, the situation is much more complex than in the Injector\,II. Those two issues are addressed for the first time in the remainder of this paper. 
\begin{table*}
\caption{\label{tab:cycs} Key parameters of the two sector cyclotrons}
\begin{tabular}{lcccccccc}
\hline \hline
& orbit radius & kin. energy &  avg. power & avg. field & peak field & magn. rigidity & harmonic number & resonators / \\
& (m) & (MeV) & (MW) & (T)& (T) & (Tm) & & cavities\\
\hline
Injector\,II & 0.44\dots3.3 & 0.87\dots72 & 0.15 & 0.33\dots0.36 & 1.08 & 1.25 & 10 & 4  \\
Ring Cyclotron & 2.1\dots4.45 & 72\dots592 & 1.3 & 0.6\dots0.9 & 2.17 & 4.0 & 6& 4 (1 flat-top)\\
\hline \hline
\end{tabular}
\end{table*}

\subsection{Different phase width studies of the PSI Ring Cyclotron}

Although a very compact beam with a phase width of about $2^\circ$ can be extracted from the Injector\,II, it is nevertheless subject to the expansion in the longitudinal direction in the 72\,MeV 
beam transfer line because of space charge effects and chromatic dispersion. For the future 3\,mA beam, this will have a significant impact on the beam dynamic of the Ring Cyclotron.
In response to this, a rebuncher running on the 10$th$ harmonic is planned to be installed on the beam line to make bunches as short as possible at the injection point of the Ring Cyclotron.
The final bunch length achieved is a critical aspect of the Ring Cyclotron.

In order to obtain a clear perspective on this issue, \opalcycl \  was applied to do numerical simulation by tracking Gaussian type beams with 3
different initial conditions. The initial longitudinal phase widths (6$\sigma$) are set to $2^\circ$, $6^\circ$ and $10^\circ$, respectively,
and the initial energy spread is neglected.
The initial conditions of the horizontal and vertical directions are identical. 
The initial distribution is not correlated in phase space.
The simulation used $10^6$ macro-particles and 32$\times$32$\times$32 gird sizes. The peak voltage of the four main resonators and the 3rd harmonic flat-top resonator are 0.9\,MV and
0.403\,MV (11.2\% of accelerating voltage), respectively. The time step is set to 0.1 ns. It takes about 7 hours on CRAY XT3 of CSCS using 64 processors to track particles from the 
injection to the extraction.

\begin{figure*}
  \includegraphics[width=0.45\linewidth]{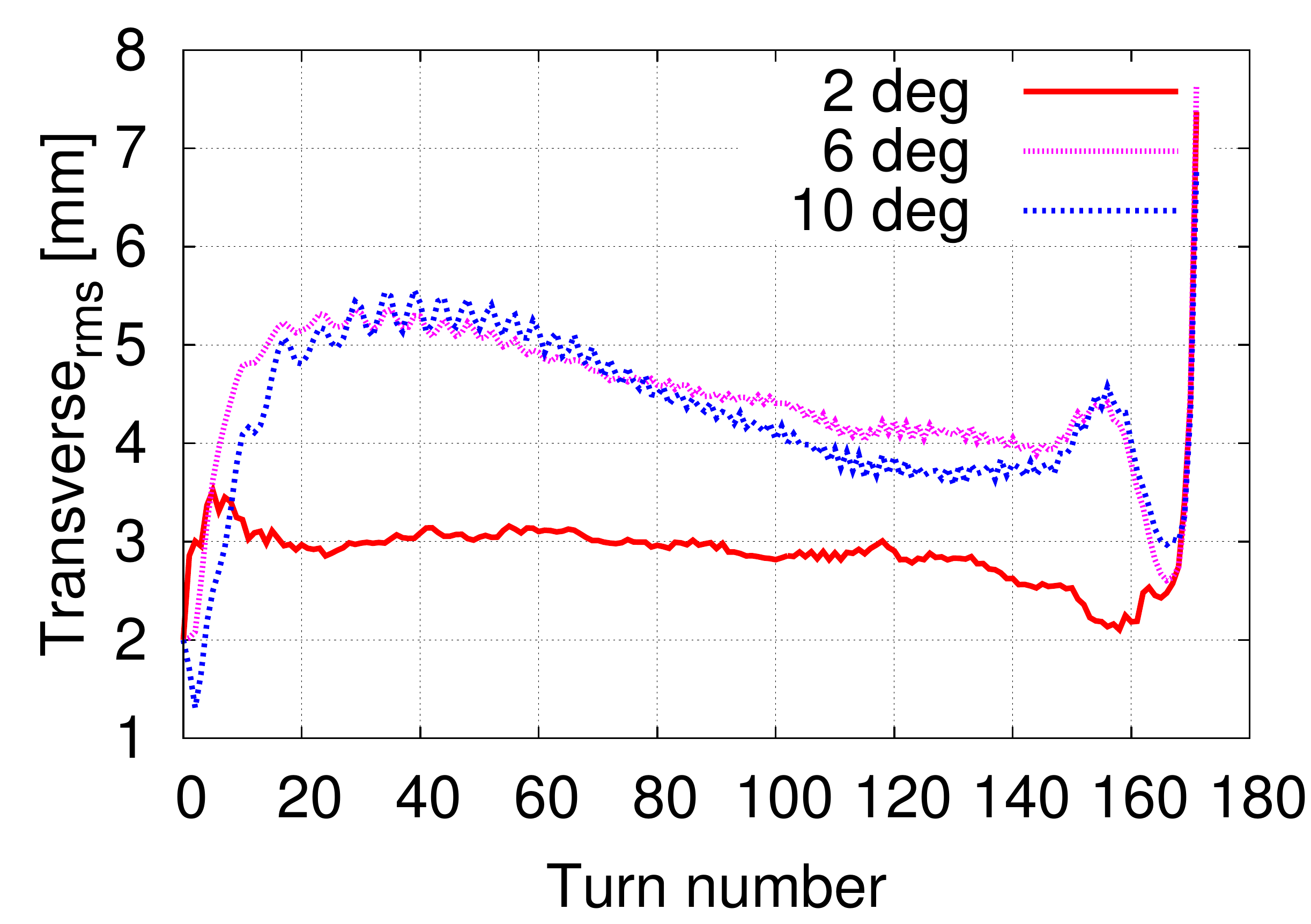}
  \includegraphics[width=0.45\linewidth]{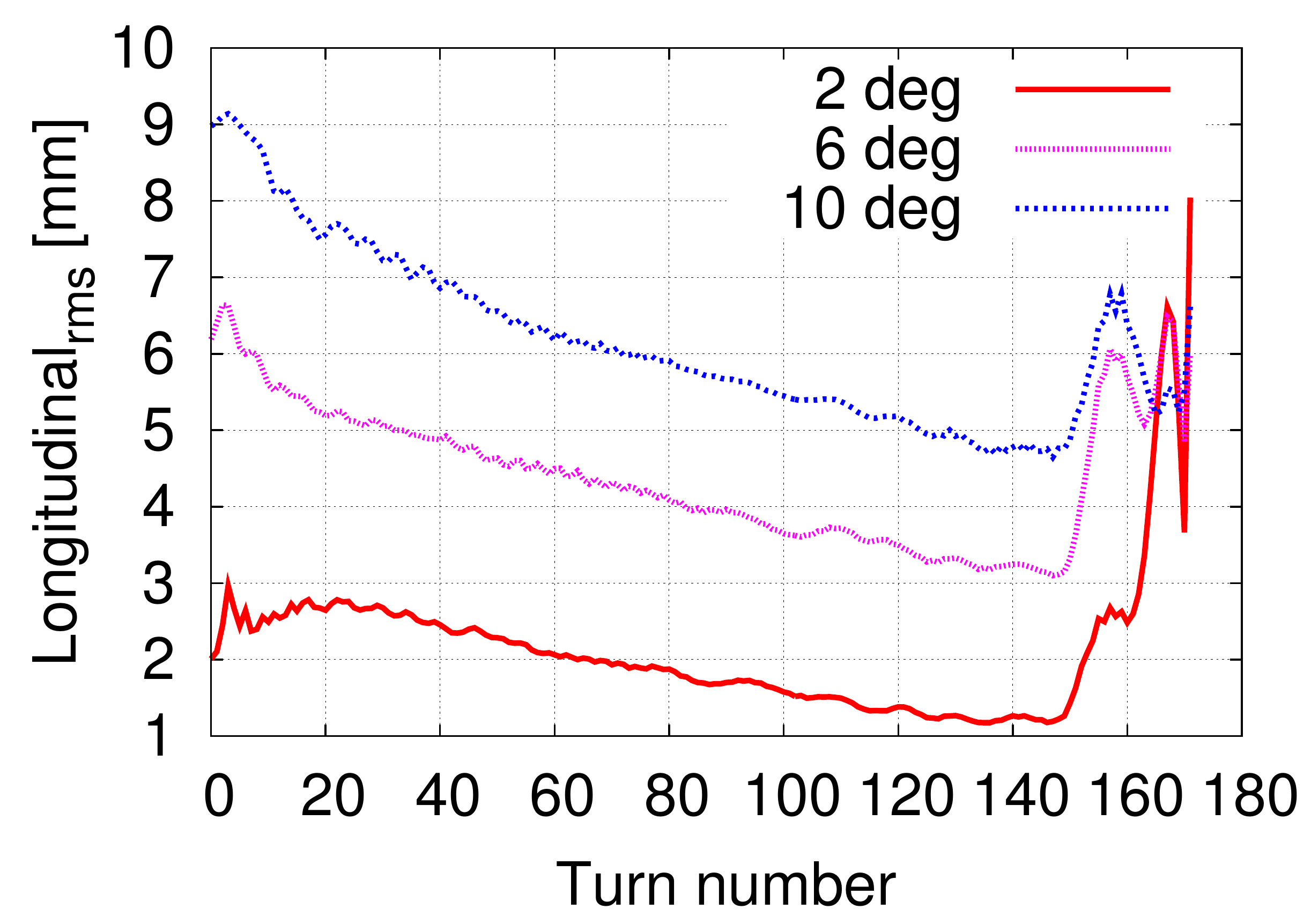}
  \caption{(Color) Comparison of the rms beam size in the transverse direction (left) and longitudinal direction (right) at $112^\circ$ azimuthal position of each turn 
    in PSI Ring Cyclotron. }
  \label{fig:RMSsize}
\end{figure*}

\begin{figure*}
  \includegraphics[angle=90,width=0.9\linewidth]{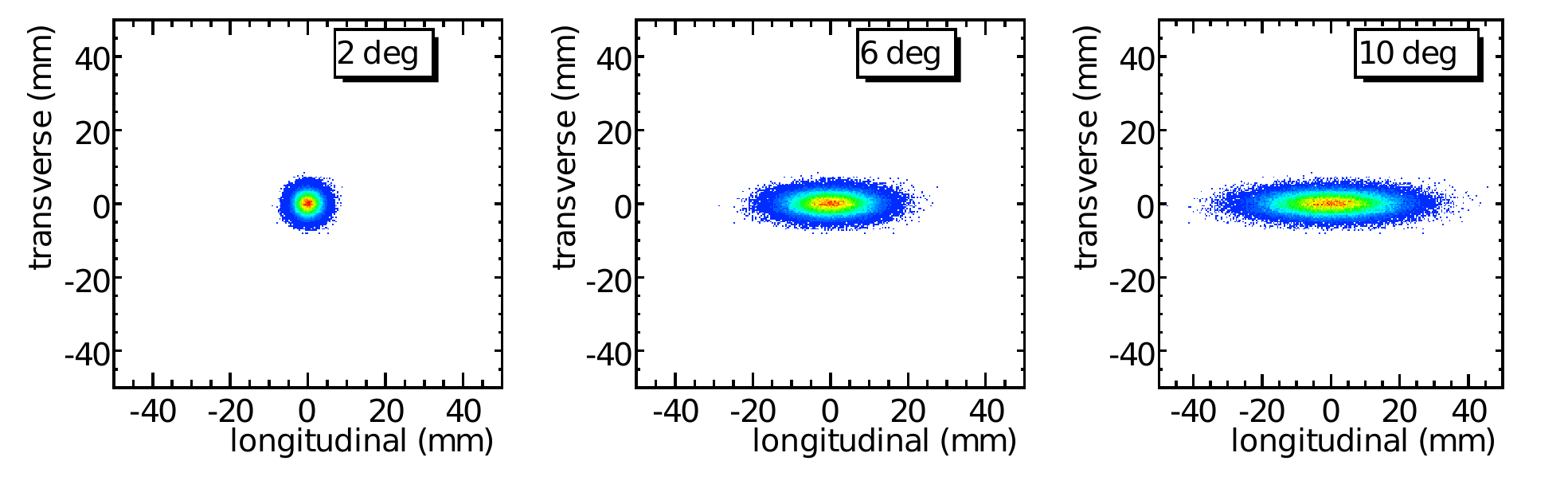}
  \includegraphics[angle=90,width=0.9\linewidth]{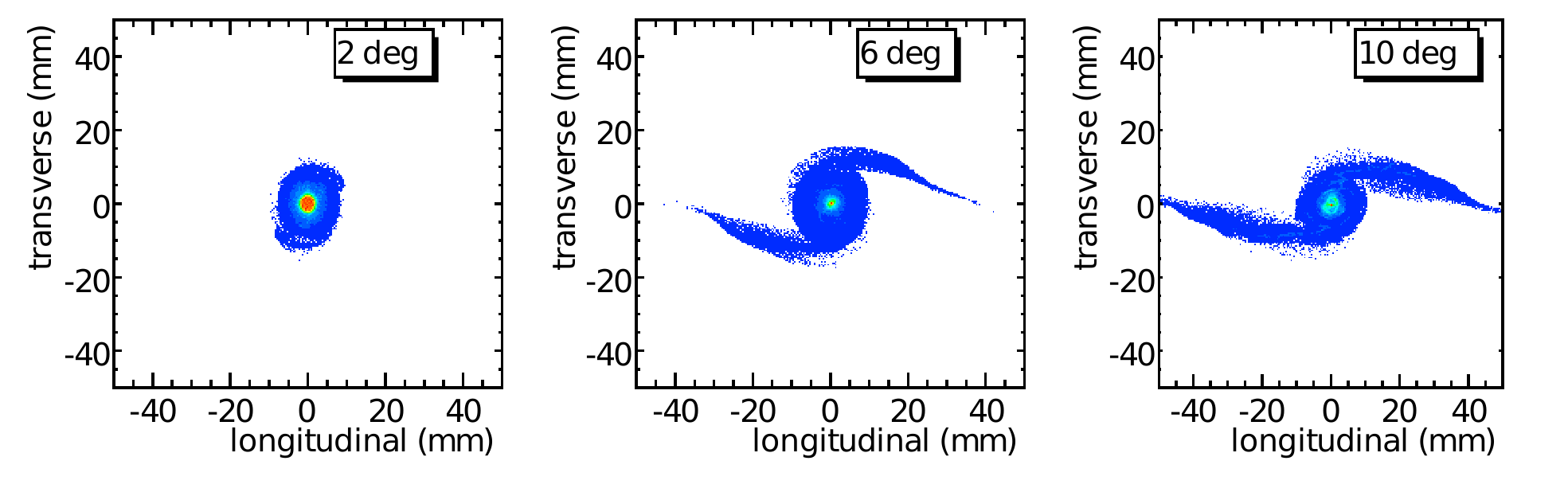}
  \includegraphics[angle=90,width=0.9\linewidth]{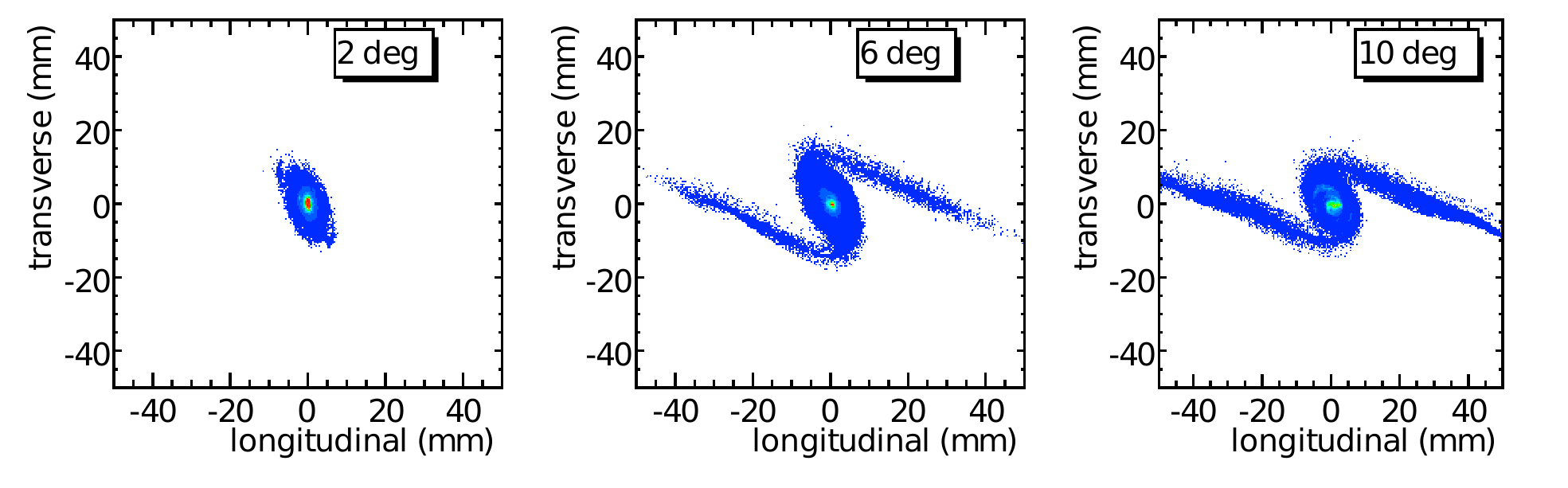}
  \caption{(Color) Top view of 3\,mA bunch distributions with  $2^\circ$, $6^\circ$ and $10^\circ$ initial phase widths at the initial position(top) turn 50 (middle), and 150 (bottom) 
    in the local frame ${\bs{S}_{local}}$ of $112^\circ$ azimuthal position of PSI Ring Cyclotron. }
  \label{fig:RingPhaseWidth}
\end{figure*}

\begin{figure*}
  \includegraphics[angle=90,width=0.3\linewidth]{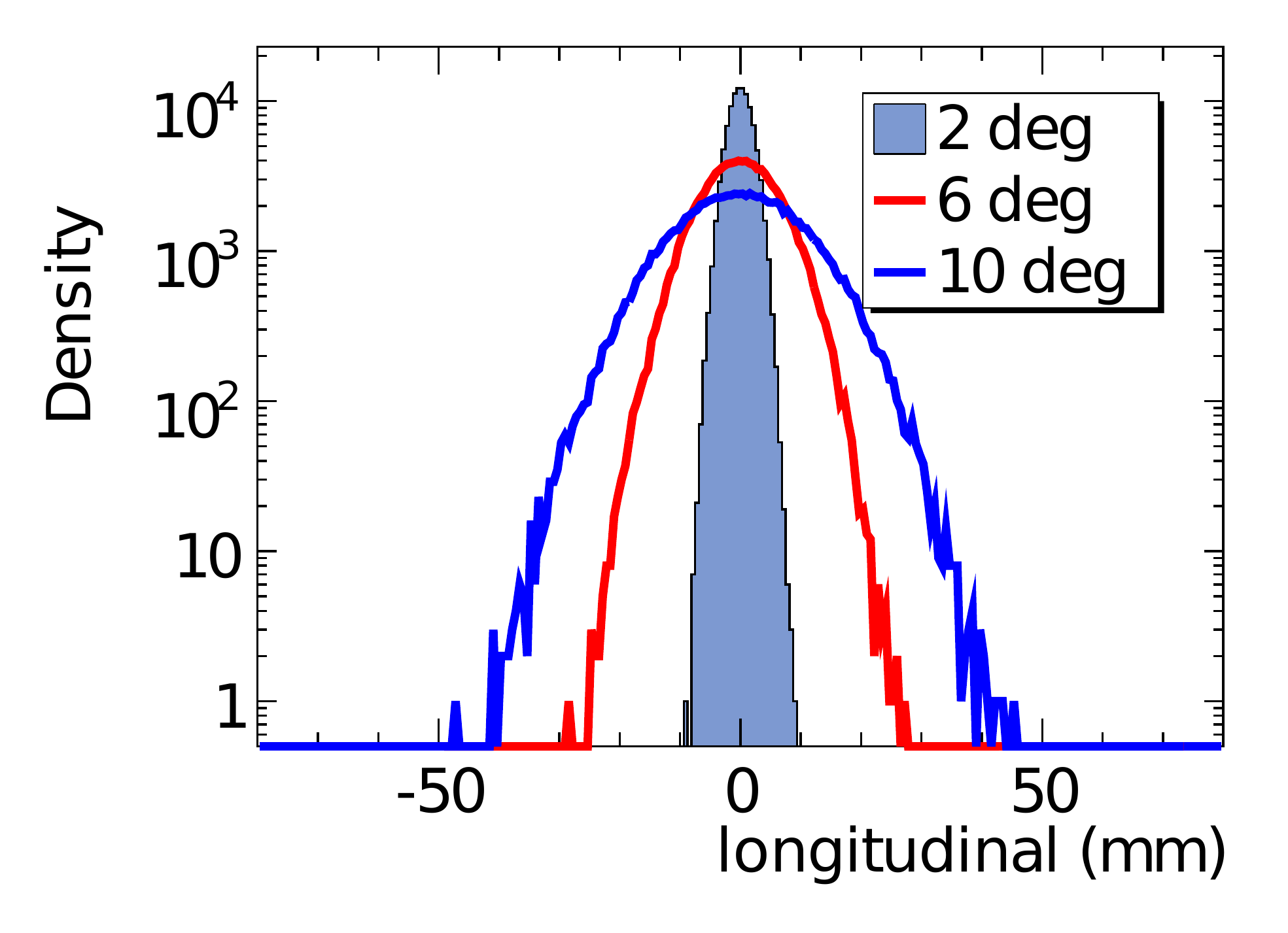}
  \includegraphics[angle=90,width=0.3\linewidth]{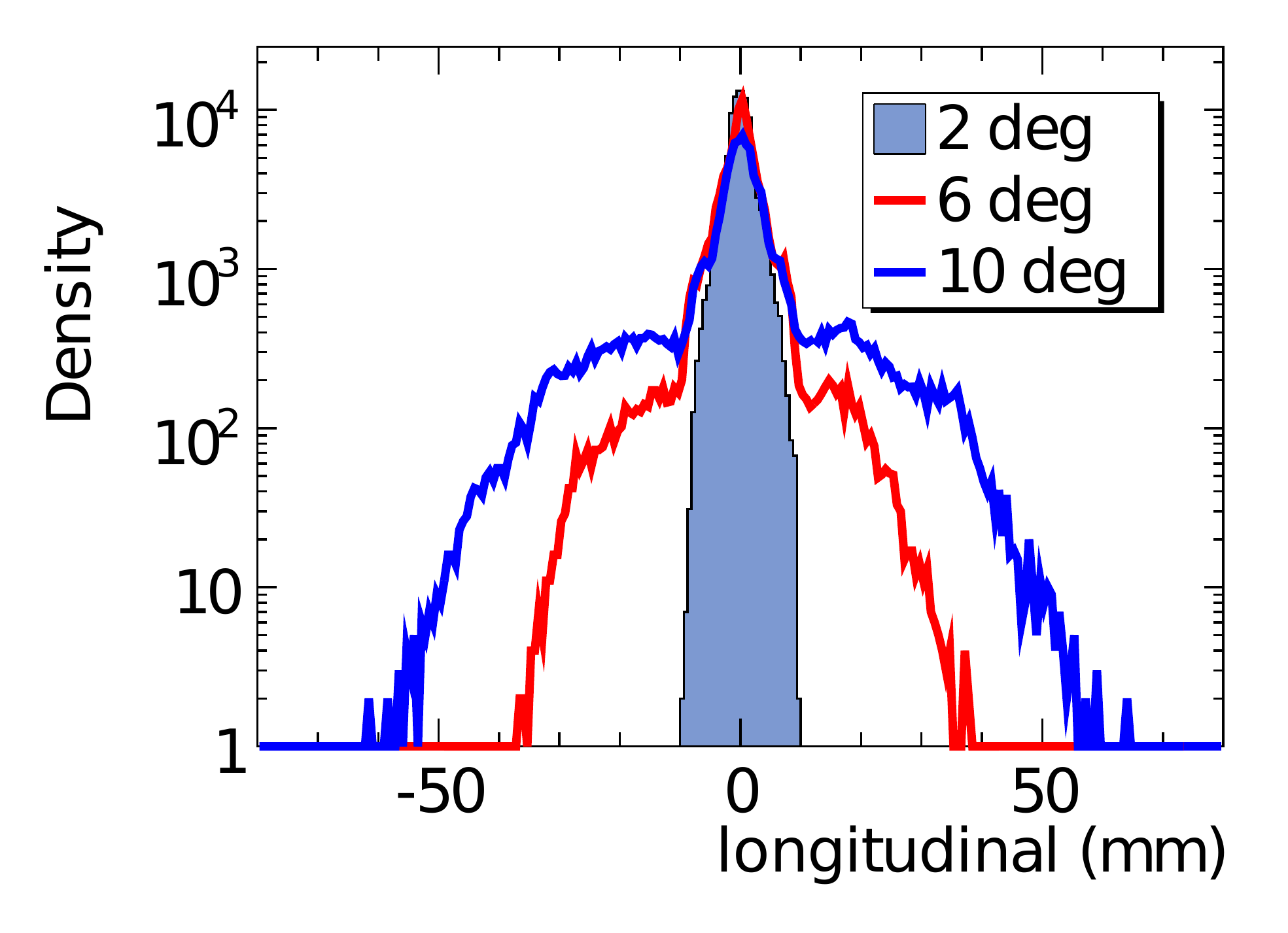}
  \includegraphics[angle=90,width=0.3\linewidth]{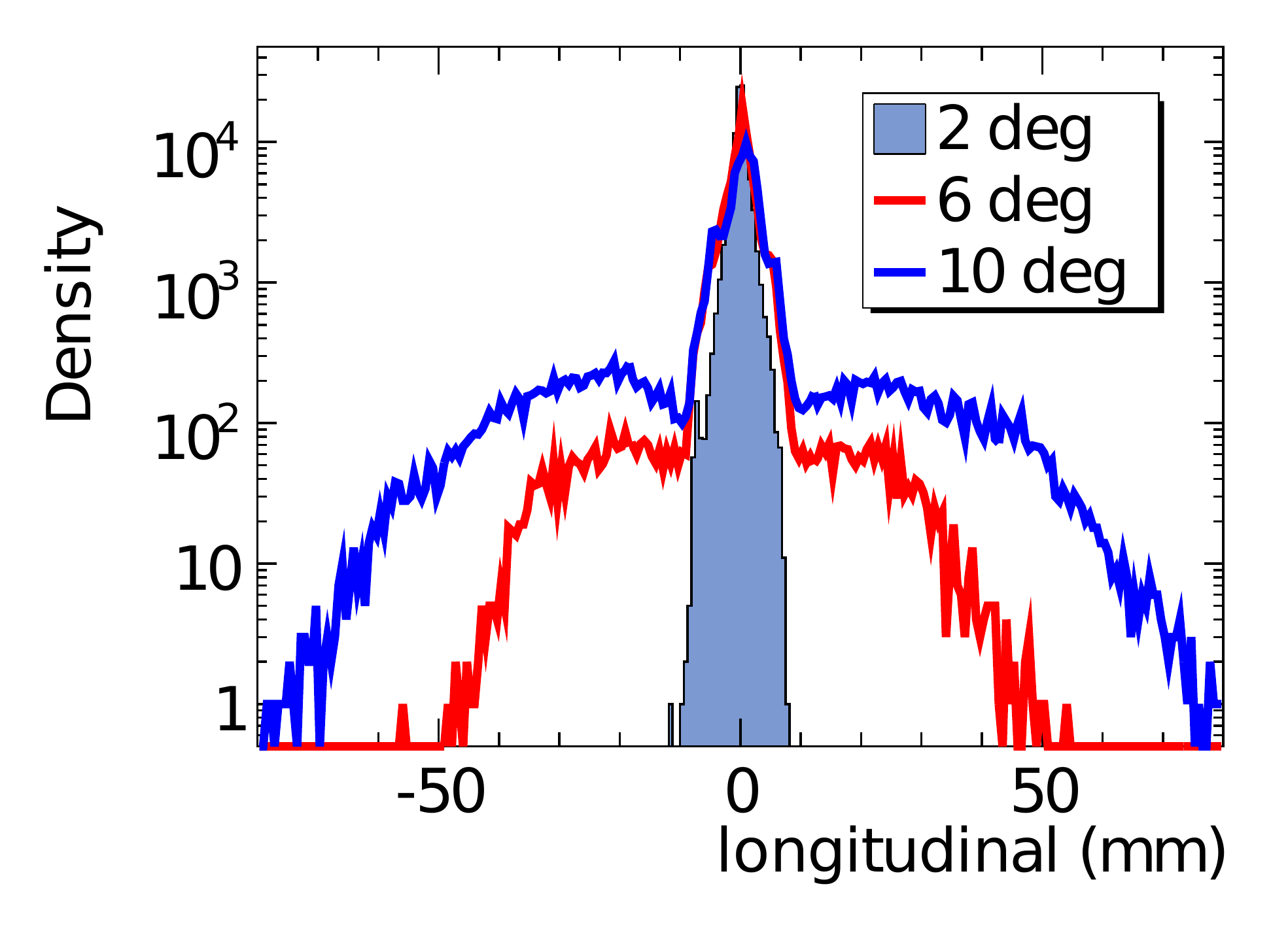}
  \caption{(Color) Histogram of 3\,mA bunch distributions with  $2^\circ$, $6^\circ$ and $10^\circ$ initial phase widths at initial position(left), turn 50 (middle), and 150 (right) in PSI Ring Cyclotron.}
  \label{fig:ThetaHitgram}
\end{figure*}

Figure\,\ref{fig:RMSsize} shows the development of the beam rms size on the transverse and the longitudinal direction. We can see the beam is compressed
gradually in the longitudinal direction. Meanwhile, in the transverse direction, the beam size increases fast during the first several turns 
because of the mismatch of initial conditions. Thereafter the beam size  does not change significantly until the beam arrives at the extraction region 
where it is distorted by the external magnetic field (extraction bump).
Figure\,\ref{fig:RingPhaseWidth} shows the projection of phase space onto the mid plane of the machine, and 
Fig.\,\ref{fig:ThetaHitgram} plots the histogram along the longitudinal direction at $112^\circ$ azimuthal position of turn 0, 50 and 150.
We can see for the bunch with the initial phase width of $2^\circ$, the bunch maintains a very compact shape with a stable round core without haloes. 
When the initial phase width increases, the size of the core only widens slightly (less than 5\,mm), while the spiral tails expand in the longitudinal direction and 
are unable to develop stable haloes. However, the beam does not expand notably in the radial direction, which means no substantial increase of the beam loss 
on the extraction septum is expected for bunches with initial phase width less than $10^\circ$.

\subsection {Neighboring bunch effects in the PSI Ring}

\begin{figure*}
  \includegraphics[width=1\linewidth]{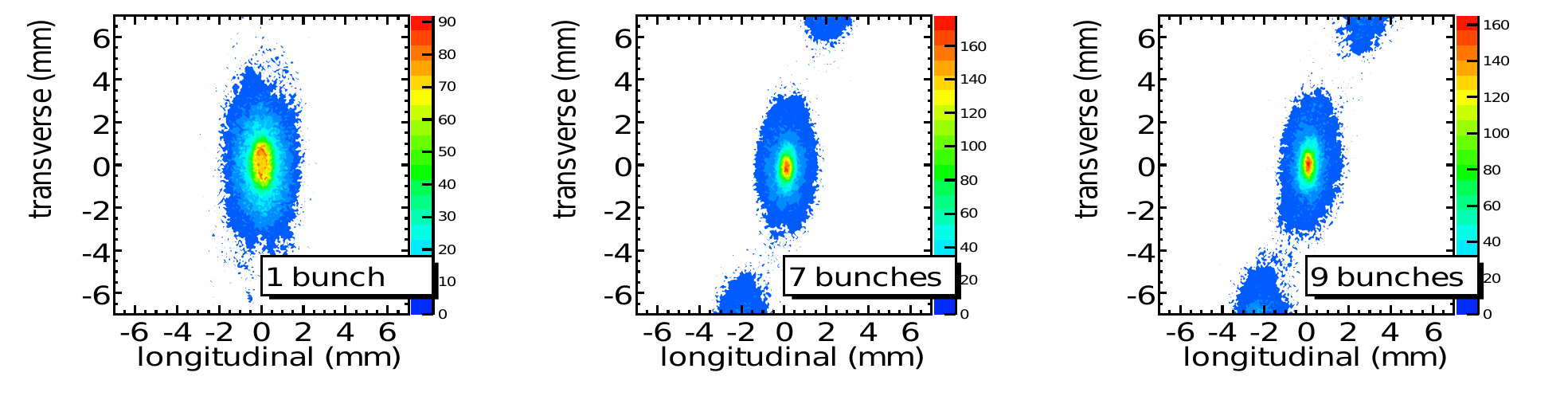}
  \caption{(Color) Top view of 1\,mA bunch distributions at the turn 130  in the local frame ${\bs{S}_{local}}$ at $112^\circ$ azimuthal position of turn 130 in PSI Ring Cyclotron.
    The results are obtained from single bunch (left), 7 bunches (middle) and 9 bunches (right) simulations, respectively.}
  \label{fig:NBcompare2D}
\end{figure*}

\begin{figure*}
  \includegraphics[width=0.45\linewidth]{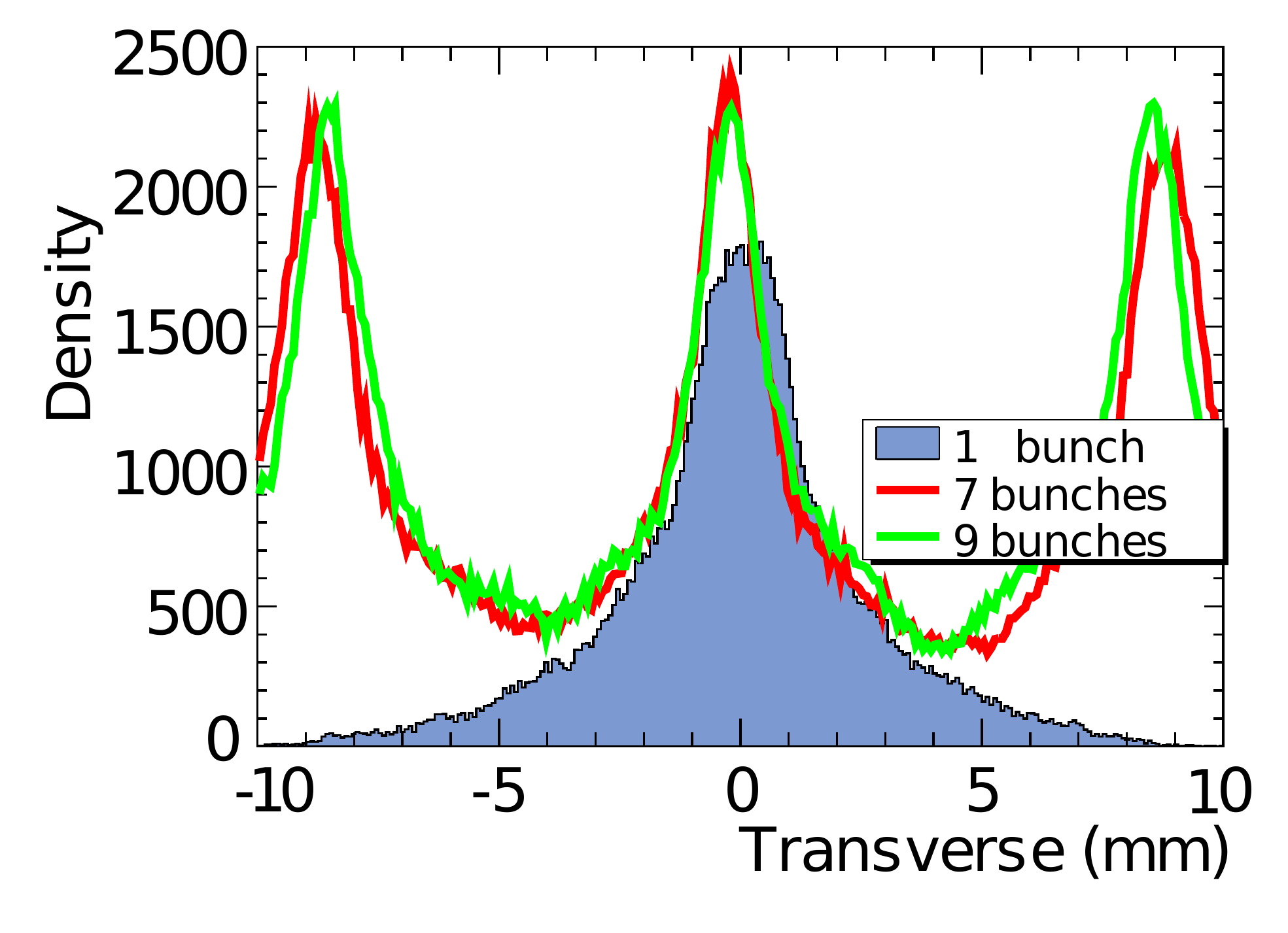}
  \includegraphics[angle=90,width=0.45\linewidth]{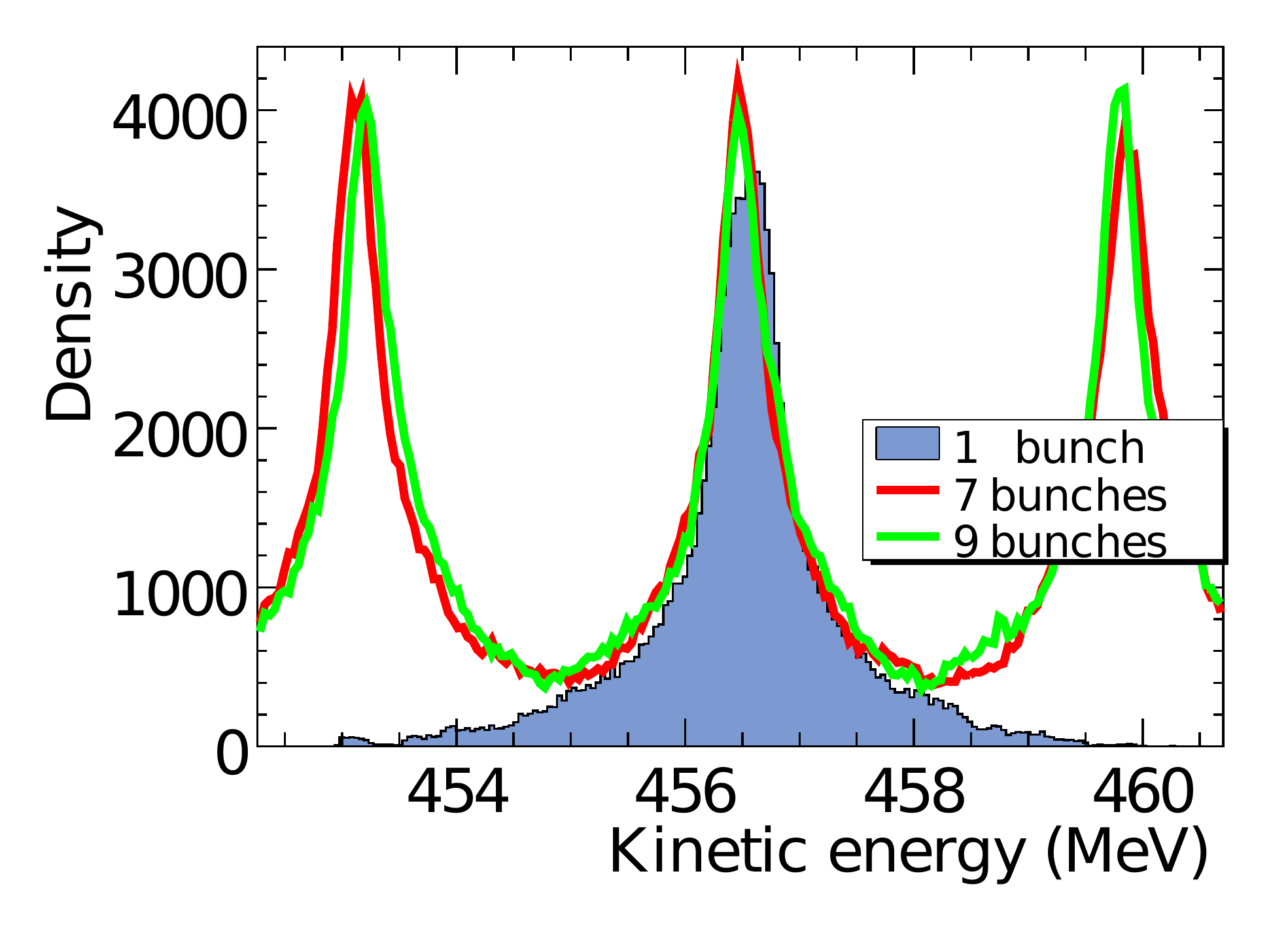}
  \caption{(Color) Comparison of the histograms along the transversal direction in the local frame ${\bs{S}_{local}}$ (left) and the energy spectra (right) of 1\,mA beam
    at $112^\circ$ azimuthal position of turn 130 in PSI Ring Cyclotron.}
  \label{fig:NBcompare}
\end{figure*}

As discussed in Section I, neighboring bunch effects may have an appreciable influence on beam dynamics in the Ring. This can be evaluated by comparing the difference in single bunch and multiple bunch simulations as 
described in Section III. We have run simulations for  3, 5, 7 and 9 bunches and found that 
the difference between the 7 bunch scenario and that of 9 bunch scenarios is small, i.e. is viewed as converged,  as illustrated in Fig.\,\ref{fig:NBcompare2D} and discussed already in Section \ref{sec:nbe}.
From Fig.\,\ref{fig:NBcompare} we conclude that the FWHM of the  transverse profile is reduced by approximately 33\% comparing a 1 and 9 bunch simulation. For the energy spread (FWHM) we have a reduction in the order
of 14\% i.e from 0.7 MeV to 0.6 MeV.

As Gordon explained in \cite{Gordon:1} the particle motion in a cyclotron is always perpendicular to the force
resulting in a vortex motion. In order to obtain the observed sharpening of the distribution we
need an additional, azimuthal force. A possible explanation of the origin of this force is due to the observed broken circular symmetry when considering neighboring bunches in the simulation. However more  efforts are needed in order to understand this effect in greater detail.


From the comparison we conclude that the integration of neighboring bunch effects into the model has non-negligible impacts on the beam dynamics for beam currents beyond 1\,mA in the PSI Ring. 
The bunch becomes more compact in the transverse direction and the energy spread is slightly reduced. 
Therefore neighboring bunch effects have a positive influence on reducing beam loss in high intensity operation.

\section{CONCLUSIONS AND DISCUSSIONS}
A physical model for the beam dynamics in high intensity cyclotrons, which includes for the first time the space charge effects
of neighboring bunches, is presented in this paper. 
This model is implemented in an object-oriented three-dimensional parallel PIC code (\opalcycl), 
as a flavor of the \opal\, framework. 

The performance tests on the CRAY XT3, CSCS demonstrate a good scalability of \opalcycl \ with respect to the number of used processors. 
The three operation modes of this code (tune calculation, single and multiple particle mode) are validated by code comparison. 
\opalcycl \ has been successfully applied to study the behavior of the PSI Ring Cyclotron at high intensities.

The beam intensity of this high power facility is practically limited by uncontrolled losses at the extraction element of the cyclotron, originating from beam tails that are developed during the acceleration process.
As shown in the results, the generation of beam tails can be avoided if short bunches with a phase length of $2^\circ$  or less are injected. 
An upgrade plan is under way to generate such short bunches with the help of a 10$th$ harmonic buncher.
Furthermore it is observed that the neighboring bunch effects can help to narrow the transverse beam size and reduce the energy spread. The differences between single and multi-bunch simulations
are in the order of 33\% and 15\% in the compared quantities, beam size and energy spread respectively. This is a significant difference between single and multi-bunch simulation, and hence justifies the presented simulation method and their application.
This is an important step towards
the quantitative understanding of beam tails in high power cyclotrons. Considering the fact that in the PSI Ring Cyclotron the total sum of controlled and uncontrolled losses are
in the order of $10^{-4}$, a precise beam dynamics simulation must cover radial neighboring tuns in order to predict losses with the mentioned intensities. 

It is planned to refine these simulations within the next few years by a more detailed determination of the initial particle distribution at the injection
of the PSI Ring Cyclotron. A quantitative comparison of the results with measured beam quantities will be presented in a future paper.
\section{ACKNOWLEDGMENTS}
The authors would like to extend sincere thanks the Accelerator Modeling and Advanced Computing group members C.\,Kraus, Y.\,Ineichen and B.\,Oswald for many useful discussions regarding programming and T.\,Schietinger for providing the post-processing tool
H5PartRoot. We also feel indebted to  W.\,Joho, S.\,Adam and R.\,D\"olling for many helpful discussions regarding high
intensity beam dynamics in cyclotrons. This work was performed on the Merlin3 cluster at the Paul Scherrer Institut 
and on the Cray XT3 at Swiss National Supercomputing Center (CSCS) within the ``Horizon'' collaboration. One author (J.\,J.\,Yang) 
was partially supported by Natural Science Foundation of China (10775185) during his sabbatical at PSI.
The authors would also like to thank the referees for a number of suggestions that have substantially improved the form of this work.

\bibliography{Refbase}

\begin{thebibliography}{31}
\expandafter\ifx\csname natexlab\endcsname\relax\def\natexlab#1{#1}\fi
\expandafter\ifx\csname bibnamefont\endcsname\relax
  \def\bibnamefont#1{#1}\fi
\expandafter\ifx\csname bibfnamefont\endcsname\relax
  \def\bibfnamefont#1{#1}\fi
\expandafter\ifx\csname citenamefont\endcsname\relax
  \def\citenamefont#1{#1}\fi
\expandafter\ifx\csname url\endcsname\relax
  \def\url#1{\texttt{#1}}\fi
\expandafter\ifx\csname urlprefix\endcsname\relax\def\urlprefix{URL }\fi
\providecommand{\bibinfo}[2]{#2}
\providecommand{\eprint}[2][]{\url{#2}}

\bibitem[{\citenamefont{Baartman}(1995)}]{Baartman:1}
\bibinfo{author}{\bibfnamefont{R.}~\bibnamefont{Baartman}}, in
  \emph{\bibinfo{booktitle}{Proc. 14th Int. Conf. on Cyclotrons and their
  Applications}} (\bibinfo{address}{Capetown}, \bibinfo{year}{1995}), p.
  \bibinfo{pages}{440}.

\bibitem[{\citenamefont{Adam}(1985{\natexlab{a}})}]{Adam:0}
\bibinfo{author}{\bibfnamefont{S.}~\bibnamefont{Adam}}, Ph.D. thesis,
  \bibinfo{school}{ETHZ, Switzerland} (\bibinfo{year}{1985}{\natexlab{a}}),
  \bibinfo{note}{no 7694}.

\bibitem[{\citenamefont{Adam}(1985{\natexlab{b}})}]{Adam:1}
\bibinfo{author}{\bibfnamefont{S.}~\bibnamefont{Adam}}, \bibinfo{journal}{IEEE
  Trans. on Nuclear Science} \textbf{\bibinfo{volume}{32}},
  \bibinfo{pages}{2507} (\bibinfo{year}{1985}{\natexlab{b}}).

\bibitem[{\citenamefont{Koscielniak and Adam}(1993)}]{Adam:2}
\bibinfo{author}{\bibfnamefont{S.}~\bibnamefont{Koscielniak}} \bibnamefont{and}
  \bibinfo{author}{\bibfnamefont{S.}~\bibnamefont{Adam}}, in
  \emph{\bibinfo{booktitle}{Proc. Particle Accelerator Conf.}}
  (\bibinfo{address}{Washington}, \bibinfo{year}{1993}), p.
  \bibinfo{pages}{3639}.

\bibitem[{\citenamefont{Bertrand and Ricaud}(2001)}]{Bert:2001}
\bibinfo{author}{\bibfnamefont{P.}~\bibnamefont{Bertrand}} \bibnamefont{and}
  \bibinfo{author}{\bibfnamefont{C.}~\bibnamefont{Ricaud}}, in
  \emph{\bibinfo{booktitle}{Conf. on Cyclotrons and their Applications}}
  (\bibinfo{address}{East Lansing, Michigan}, \bibinfo{year}{2001}), p.
  \bibinfo{pages}{379}.

\bibitem[{\citenamefont{Adelmann}(2002)}]{Ada:1}
\bibinfo{author}{\bibfnamefont{A.}~\bibnamefont{Adelmann}}, Ph.D. thesis,
  \bibinfo{school}{ETHZ, Switzerland} (\bibinfo{year}{2002}), \bibinfo{note}{no
  14545}.

\bibitem[{\citenamefont{Pozdeyev}(2003)}]{Poz:1}
\bibinfo{author}{\bibfnamefont{E.}~\bibnamefont{Pozdeyev}}, Ph.D. thesis,
  \bibinfo{school}{MSU, USA} (\bibinfo{year}{2003}).

\bibitem[{\citenamefont{Humbel}(2009)}]{HumbPC}
\bibinfo{author}{\bibfnamefont{M.}~\bibnamefont{Humbel}},
  \bibinfo{howpublished}{private communication} (\bibinfo{year}{2009}).

\bibitem[{\citenamefont{Kleeven}(1988)}]{kleeven:1}
\bibinfo{author}{\bibfnamefont{W.}~\bibnamefont{Kleeven}}, Ph.D. thesis,
  \bibinfo{school}{TU Eindhoven} (\bibinfo{year}{1988}).

\bibitem[{\citenamefont{Gordon}(1969)}]{Gordon:1}
\bibinfo{author}{\bibfnamefont{M.~M.} \bibnamefont{Gordon}}, in
  \emph{\bibinfo{booktitle}{Proc. 5th Int. Conf. on Cyclotrons and their
  Applications}} (\bibinfo{address}{Oxford}, \bibinfo{year}{1969}), p.
  \bibinfo{pages}{305}.

\bibitem[{\citenamefont{Joho}(1981)}]{Joho:1}
\bibinfo{author}{\bibfnamefont{W.}~\bibnamefont{Joho}}, in
  \emph{\bibinfo{booktitle}{Proc. 9th Int. Conf. on Cyclotrons and their
  Applications}} (\bibinfo{address}{Caen}, \bibinfo{year}{1981}), p.
  \bibinfo{pages}{337}.

\bibitem[{\citenamefont{Li}(2001)}]{Li:1}
\bibinfo{author}{\bibfnamefont{H.}~\bibnamefont{Li}}, Ph.D. thesis,
  \bibinfo{school}{GUCAS, China} (\bibinfo{year}{2001}).

\bibitem[{\citenamefont{Hockney and Eastwood}(1988)}]{Hockney:1}
\bibinfo{author}{\bibfnamefont{R.~W.} \bibnamefont{Hockney}} \bibnamefont{and}
  \bibinfo{author}{\bibfnamefont{J.~W.} \bibnamefont{Eastwood}},
  \emph{\bibinfo{title}{Computer Simulation Using Particles}}
  (\bibinfo{publisher}{Hilger}, \bibinfo{address}{New York},
  \bibinfo{year}{1988}).

\bibitem[{\citenamefont{Qiang et~al.}(2000)\citenamefont{Qiang, Ryne, and
  et~al.}}]{Qiang:1}
\bibinfo{author}{\bibfnamefont{J.}~\bibnamefont{Qiang}},
  \bibinfo{author}{\bibfnamefont{R.}~\bibnamefont{Ryne}}, \bibnamefont{and}
  \bibinfo{author}{\bibfnamefont{S.~H.} \bibnamefont{et~al.}},
  \bibinfo{journal}{J. Comput. Phys.} \textbf{\bibinfo{volume}{163}},
  \bibinfo{pages}{434} (\bibinfo{year}{2000}).

\bibitem[{\citenamefont{Galambos et~al.}(1999)\citenamefont{Galambos, Danilov,
  and et~al.}}]{Gala:1}
\bibinfo{author}{\bibfnamefont{J.}~\bibnamefont{Galambos}},
  \bibinfo{author}{\bibfnamefont{S.}~\bibnamefont{Danilov}}, \bibnamefont{and}
  \bibinfo{author}{\bibfnamefont{D.~J.} \bibnamefont{et~al.}}, in
  \emph{\bibinfo{booktitle}{Proc. Particle Accelerator Conf.}}
  (\bibinfo{address}{New York}, \bibinfo{year}{1999}), p.
  \bibinfo{pages}{3143}.

\bibitem[{\citenamefont{Grote and et~al.}(1996)}]{Grote:1}
\bibinfo{author}{\bibfnamefont{D.}~\bibnamefont{Grote}} \bibnamefont{and}
  \bibinfo{author}{\bibfnamefont{A.~F.} \bibnamefont{et~al.}},
  \bibinfo{journal}{Fusion Engineering \& Design}
  \textbf{\bibinfo{volume}{32}}, \bibinfo{pages}{193} (\bibinfo{year}{1996}).

\bibitem[{\citenamefont{Huang and et~al.}(2006)}]{Huang:1}
\bibinfo{author}{\bibfnamefont{C.}~\bibnamefont{Huang}} \bibnamefont{and}
  \bibinfo{author}{\bibfnamefont{V.~K.~D.} \bibnamefont{et~al.}},
  \bibinfo{journal}{J. Comput. Phys.} \textbf{\bibinfo{volume}{217}},
  \bibinfo{pages}{658} (\bibinfo{year}{2006}).

\bibitem[{\citenamefont{Seidel and Schmelzbach}(2007)}]{Mike:1}
\bibinfo{author}{\bibfnamefont{M.}~\bibnamefont{Seidel}} \bibnamefont{and}
  \bibinfo{author}{\bibfnamefont{P.}~\bibnamefont{Schmelzbach}}, in
  \emph{\bibinfo{booktitle}{Proc. 18th Int. Conf. on Cyclotrons and their
  Applications}} (\bibinfo{address}{Catania}, \bibinfo{year}{2007}), p.
  \bibinfo{pages}{157}.

\bibitem[{\citenamefont{Zhang et~al.}(2007)\citenamefont{Zhang, Li, and
  Chu}}]{Zhang:1}
\bibinfo{author}{\bibfnamefont{T.}~\bibnamefont{Zhang}},
  \bibinfo{author}{\bibfnamefont{Z.}~\bibnamefont{Li}}, \bibnamefont{and}
  \bibinfo{author}{\bibfnamefont{C.}~\bibnamefont{Chu}}, in
  \emph{\bibinfo{booktitle}{Proc. 18th Int. Conf. on Cyclotrons and their
  Applications}} (\bibinfo{address}{Catania}, \bibinfo{year}{2007}),
  p.~\bibinfo{pages}{33}.

\bibitem[{\citenamefont{Gordon and Taivassalo}(1985)}]{Gordon:2}
\bibinfo{author}{\bibfnamefont{M.~M.} \bibnamefont{Gordon}} \bibnamefont{and}
  \bibinfo{author}{\bibfnamefont{V.}~\bibnamefont{Taivassalo}},
  \bibinfo{journal}{IEEE Trans. Nucl. Sci.} \textbf{\bibinfo{volume}{32}},
  \bibinfo{pages}{2447} (\bibinfo{year}{1985}).

\bibitem[{\citenamefont{M.Bassetti and G.A.Erskine}(1980)}]{bassersk}
\bibinfo{author}{\bibnamefont{M.Bassetti}} \bibnamefont{and}
  \bibinfo{author}{\bibnamefont{G.A.Erskine}}, \bibinfo{type}{Tech. Rep.}
  \bibinfo{number}{CERN-ISR-TH/80-06}, \bibinfo{institution}{CERN}
  (\bibinfo{year}{1980}).

\bibitem[{\citenamefont{Fubiani et~al.}(2006)\citenamefont{Fubiani, Qiang,
  Esarey, and Leemans}}]{Fubiani:2006p305}
\bibinfo{author}{\bibfnamefont{G.}~\bibnamefont{Fubiani}},
  \bibinfo{author}{\bibfnamefont{J.}~\bibnamefont{Qiang}},
  \bibinfo{author}{\bibfnamefont{E.}~\bibnamefont{Esarey}}, \bibnamefont{and}
  \bibinfo{author}{\bibfnamefont{W.~P.} \bibnamefont{Leemans}},
  \bibinfo{journal}{Phys. Rev. ST Accel. Beams} \textbf{\bibinfo{volume}{9}},
  \bibinfo{pages}{1} (\bibinfo{year}{2006}).

\bibitem[{\citenamefont{Adelmann et~al.}(2008)\citenamefont{Adelmann, Kraus,
  Ineichen, Russel, and Yang}}]{opal:1}
\bibinfo{author}{\bibfnamefont{A.}~\bibnamefont{Adelmann}},
  \bibinfo{author}{\bibfnamefont{C.}~\bibnamefont{Kraus}},
  \bibinfo{author}{\bibfnamefont{Y.}~\bibnamefont{Ineichen}},
  \bibinfo{author}{\bibfnamefont{S.}~\bibnamefont{Russel}}, \bibnamefont{and}
  \bibinfo{author}{\bibfnamefont{J.}~\bibnamefont{Yang}}, \bibinfo{type}{Tech.
  Rep.} \bibinfo{number}{PSI-PR-08-02}, \bibinfo{institution}{Paul Scherrer
  Institut} (\bibinfo{year}{2008}).

\bibitem[{\citenamefont{Iselin}(1996)}]{Classic:1}
\bibinfo{author}{\bibfnamefont{F.}~\bibnamefont{Iselin}}, in
  \emph{\bibinfo{booktitle}{Comput. Accelerator Physics Conf.}}
  (\bibinfo{address}{Williamsburg}, \bibinfo{year}{1996}).

\bibitem[{\citenamefont{Adelmann}(2009)}]{ippl:1}
\bibinfo{author}{\bibfnamefont{A.}~\bibnamefont{Adelmann}},
  \bibinfo{type}{Tech. Rep.} \bibinfo{number}{PSI-PR-09-05},
  \bibinfo{institution}{Paul Scherrer Institut} (\bibinfo{year}{2009}).

\bibitem[{\citenamefont{Adelmann et~al.}(2005)\citenamefont{Adelmann, Ryne,
  Shalf, and Siegerist}}]{H5part:1}
\bibinfo{author}{\bibfnamefont{A.}~\bibnamefont{Adelmann}},
  \bibinfo{author}{\bibfnamefont{R.}~\bibnamefont{Ryne}},
  \bibinfo{author}{\bibfnamefont{J.}~\bibnamefont{Shalf}}, \bibnamefont{and}
  \bibinfo{author}{\bibfnamefont{C.}~\bibnamefont{Siegerist}}, in
  \emph{\bibinfo{booktitle}{Proc. Particle Accelerator Conf.}}
  (\bibinfo{address}{Knoxville}, \bibinfo{year}{2005}), p.
  \bibinfo{pages}{4129}.

\bibitem[{\citenamefont{Schietinger}(2009)}]{h5partroot:1}
\bibinfo{author}{\bibfnamefont{T.}~\bibnamefont{Schietinger}},
  \emph{\bibinfo{title}{H5PartROOT: a ROOT Based Graphical User Interface for
  H5Part}},
  \bibinfo{organization}{\url{http://amas.web.psi.ch/tools/H5PartROOT/index.ht%
ml}} (\bibinfo{year}{2009}).

\bibitem[{\citenamefont{Rudolf}(2000)}]{FIXPO:1}
\bibinfo{author}{\bibfnamefont{G.}~\bibnamefont{Rudolf}},
  \emph{\bibinfo{title}{FIXPO Bedienungsanleitung}},
  \bibinfo{organization}{PSI} (\bibinfo{year}{2000}).

\bibitem[{\citenamefont{Joho}(1970)}]{joho:2}
\bibinfo{author}{\bibfnamefont{W.}~\bibnamefont{Joho}}, \bibinfo{type}{Tech.
  Rep.} \bibinfo{number}{TM-11-07}, \bibinfo{institution}{Paul Scherrer
  Institut} (\bibinfo{year}{1970}).

\bibitem[{\citenamefont{Adam}(1995)}]{Adam:3}
\bibinfo{author}{\bibfnamefont{S.}~\bibnamefont{Adam}}, in
  \emph{\bibinfo{booktitle}{Proc. 14th Int. Conf. on Cyclotrons and their
  Applications}} (\bibinfo{address}{Cape Town}, \bibinfo{year}{1995}), p.
  \bibinfo{pages}{446}.

\bibitem[{\citenamefont{Adam}(2009)}]{adapconv}
\bibinfo{author}{\bibfnamefont{S.}~\bibnamefont{Adam}},
  \bibinfo{howpublished}{private communication on PICN model}
  (\bibinfo{year}{2009}).

\end{thebibliography}

\end{document}